# Thermal Confinement in Plasmonic Nanoparticles under Generalised Pulsed Illumination

Javier González-Colsa[a,b,*], Dolores Ortiz[b], José María Saiz[b], Alfredo Franco[b], Francisco González[b], Fernando Bresme[c], Pablo Albella[b]

[a]CINBIO, University of Vigo, Campus Universitario de Vigo, Lagoas Marcosede, Vigo 36310, Spain.

*E-mail: javier.gonzalez.colsa@uvigo.gal.

[b]Group of Optics, Department of Applied Physics, University of Cantabria, 39005 Santander, Spain.

[c]Department of Chemistry, Molecular Sciences Research Hub, Imperial College London, London W12 0BZ, UK.

**ABSTRACT**

Thermal confinement is a key parameter in photothermal applications, as it determines the spatial extent of the region affected by the generated heat. In this work, we develop a numerical treatment to quantify thermal confinement in spherical plasmonic nanoparticles under excitation conditions ranging from ultrashort pulses to the continuous-wave regime. The model combines generalized optical heating with a finite-difference solution of the transient heat equation and explicitly accounts for finite interfacial thermal conductance. To characterize the spatial distribution of heat, we introduce a thermal confinement radius defined from the temporal temperature envelope in the surrounding medium. We demonstrate that pulse duration acts as a highly effective, tuneable parameter to govern thermal confinement. Notably, we quantitatively show that finite interfacial thermal conductance can induce deviations of up to 20% in the spatial confinement radius and over 60% in the transient thermal construct under ultrashort excitation. By comparing the characteristic dimensions of representative targeting ligands with the resulting thermal-confinement length scales, we show how pulse duration can be precisely tailored to control target heating. These results provide a unified framework for predicting nanoscale heat localization paving the way for the rational design of high precision thermoplasmonic applications.

## INTRODUCTION

Plasmonic Nanoparticles (PNPs) have attracted great attention in Chemistry, Medicine and Physics due to their ability to transform optical energy into heat through the generation of Localized Surface Plasmon Resonances (LSPRs). LSPRs enable the remote control of temperature-driven phenomena in a diverse range of applications in which PNPs play a central role such as, optofluidics [1] , renewable energy [2] and nanomedicine [3,4]. The field of Thermoplasmonics is committed to the study of light-to-heat conversion in PNPs to increase their efficiency and applicability towards the applications. Among the strategies to optimize the thermal performance of PNPs, several approaches have been investigated in the literature. Exploiting the spectral tunability of LSPRs is a fundamental and widely used approach [5] . A huge set of PNPs have been proposed in recent years with different materials, sizes and geometries that allow for an enhanced thermoplasmonic response across the Vis-NIR spectrum, for instance, spheroids, cubes, bipyramids, rods, disks, stars or toroids [6–8], among others.

Although powerful, the thermal response of these structures is highly symmetric in space, thus providing a homogeneous heating of the surrounding medium. This can be an issue in applications where the spatial control of temperature is key, like in PhotoThermal Therapies (PTT) where the particle is meant to heat only the target cell [9]. An alternative strategy to improve spatial selectivity and implement thermal directionality consists in combining different materials with contrasting thermal properties within the same nanostructure giving rise to the concept of Janus Nanoparticles (JNPs) [10–12] . Several examples of these structures can be found in the literature such as bimaterial spheres [13], silica/gold core-semishells [14,15] or hybrid toroid-based nanostructures [16–18].

Despite these strategies, it is common in medicine to rely on pulsed illumination since every component of the tissue possesses a different characteristic heating time. As a result, the selectivity of tissue heating is controlled by the pulse duration. This concept can be extended to the case of a PNP. Analytic solutions in the literature successfully describe the continuous-wave (CW) and ultrafast asymptotic limits (typically fs for gold PNPs), where the temperature envelope decays following a $\propto 1/r$ and $\propto 1/r^3$ law, respectively [19], a unified theoretical framework capturing the intermediate confinement regimes is still missing. However, this suggests that, for the same maximum temperature in the system, the thermal field is significantly more confined under ultrafast excitation than under continuous-wave (CW) illumination. These contrasting spatial distributions raise the fundamental question of whether a continuous transition exists between the diffusion-dominated and the impulsive regimes.

If so, the pulse duration could be used as a control parameter to tune the spatial and temporal extent of the thermal field, enabling one to determine, for a given PNP, how far and how long heat propagates simply by adjusting the excitation conditions.

This aspect is particularly relevant in practical PTT, where PNPs are functionalized with ligands to selectively bind to cancerous cells. Both the PNP/fluid Interfacial Thermal Conductance (ITC) and the length of these ligands determine the effective separation and effect between the PNP surface and the biological target as well as the heat diffusion. Therefore, introducing pulse duration as an additional control parameter may intrinsically enhance treatment specificity, enabling a more precise localization of the temperature field.

Among the most representative ligands are antibodies, aptamers, peptides, and small molecules. Antibodies, particularly monoclonal IgG, are amongst the largest, with typical dimensions on the order of 10-15 nm [20–22] , leading to relatively large PNP-cell distance. In contrast, aptamers (short single-stranded DNA or RNA sequences) are significantly smaller, typically in the 1-7 nm range [23–25] depending on their length and secondary structure. Specific examples include PD-1-targeting aptamers (6.5 nm) and AS1411 (3.2nm) illustrating the variability within this class.

This variability is particularly relevant in the context of interfacial thermal transport, as ligand functionalization introduces an additional interfacial length scale and can strongly modify the heat-transfer pathways between the nanoparticle and the surrounding medium, potentially acting either as a thermal channel or as a barrier [26,27] . In both cases, understanding thermal confinement remains essential for optimizing thermal efficiency at subcellular scales.

In this work, we systematically investigate the role of pulse duration in shaping the thermal response of PNPs, with particular emphasis on the existence of intermediate confinement regimes between the CW and ultrafast limits. By combining analytical insights with in-house numerical modelling, we analyse how the temporal profile of the excitation influences the spatial and temporal evolution of the temperature field. Special attention is given to the interplay between pulse duration and interfacial effects, incorporating the role of ligand-induced separation as an additional geometric parameter. Our results demonstrate that pulse duration can be used as an effective control knob to tailor thermal confinement, providing a framework to predict and optimize heat localization under realistic conditions relevant to photothermal applications.

## METHODS

### 1. Calculation of Heat Power

In order to calculate the heat power, $\bar{Q}$, we focus on spherical gold NPs immersed in water and consider that they are under the quasistatic approximation (QA). In this context, $\bar{Q}$ can be considered homogeneous within the particle's volume. This simplification is justified for small PNPs, as the electromagnetic field fully penetrates the metallic volume, ensuring a uniform distribution of resistive losses [28]. Furthermore, the rapid thermalization time of the conduction electrons is negligible compared to the characteristic heating time of the aqueous environment, validating this approach. Therefore, $\bar{Q}$ can be described as:

$$\bar{Q}(r,t,\tau_0) = \begin{cases} Q(t,\tau_0) & r < a \\ 0 & r \geq a \end{cases} \tag{1}$$

where $a$ is the particle's radius. Relying on established models [29], the input power has been modelled considering both the pulsed excitation and the intrinsic electron-phonon scattering process. Consequently, a step function with four regimes have been implemented for $Q(t,\tau_0)$:

$$Q(t,\tau_0) = \begin{cases} \sigma_{abs}F/\tau_e \exp\left(-t/\tau_e\right)\Theta(t) & \text{if } \tau_0 \in [\tau_{-\infty}, 0.1\tau_e) \\ \sigma_{abs}F/\tau_{eff} \exp\left(-t/\tau_e\right) \operatorname{erfc}\left(\sqrt{\pi}\left(\frac{1}{2\pi}\frac{\tau_0}{\tau_e} - \frac{t}{\tau_0}\right)\right) & \text{if } \tau_0 \in [0.1\tau_e, 33\tau_e] \\ \sigma_{abs}F/\tau_0 \exp\left(-\pi\, t^2/\tau_0^2\right) & \text{if } \tau_0 \in (33\tau_e, \tau_{+\infty}] \\ \sigma_{abs}I & \text{if } \tau_0 \in (\tau_{+\infty}, +\infty) \end{cases} \tag{2}$$

where $\sigma_{abs}$, $F$ and $I$ are the absorption cross section, optical fluence and intensity. $\tau_e$ and $\tau_0$ are the electron-phonon scattering and the pulse duration time scales, and $\tau_{+\infty}$ and $\tau_{-\infty}$ are the maximum and minimum validity ranges for the Gaussian and exponential branches respectively. Besides, $\tau_{eff} = 2\tau_e / \exp\left(\frac{\tau_0^2}{4\pi\tau_e^2}\right)$. Finally, $\Theta(t)$ and $\mathrm{erfc(t)}$ are the Heaviside's and the complementary error functions, respectively. The rationale behind this implementation and the full mathematical description are shown in Section S1 of the supplementary document. In particular, the definition of the practical criterion followed to establish the temporal validity range $[0.1\tau_e, 33\tau_e]$, can be found in section S1(c) and in Figure S1. It is worth noting that this validity range can be easily adapted for different plasmonic materials.

### 2.A Crank-Nicholson scheme to solve the Heat Transfer Equation.

**Figure 1** shows the model system considered in this work, consisting of a spherical gold PNP of radius $a$ embedded in a homogeneous medium. The PNP is heated by the heat power density defined in equation (1). The optical properties of gold were taken from the Johnson & Cristy dataset [30] whereas the surrounding medium was chosen to reproduce the optical and thermophysical properties of water ($n = 1.33$). The corresponding mass density, specific heat and thermal conductivity of each material (gold, water) are respectively: $\rho = (19300, 1000)$ kg·m$^{-3}$, $c_p = (129, 4181)$ J·kg$^{-1}$·K$^{-1}$ and $k = (318, 0.6)$ W·m$^{-1}$·K$^{-1}$.

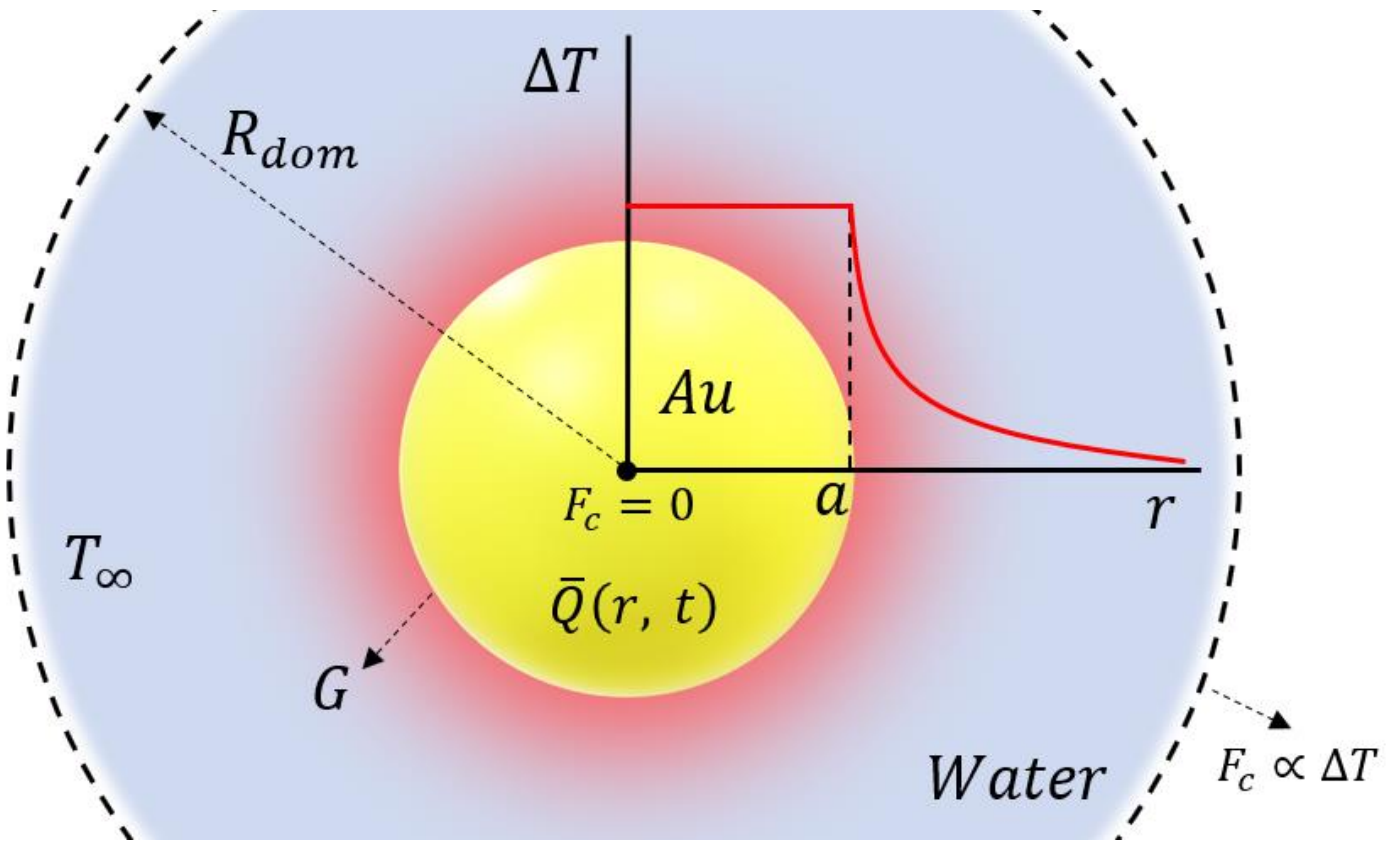

Figure 1. Representation of a spherical nanoparticle heated up by a homogeneous heat power $\bar{Q}(r,t,\tau_0)$ in equation (1). $\Delta T(r)$ represents the thermal profile obtained after illumination. $R_{dom}$, $G$ and $T_\infty$ are the size of the simulation region, the ITC and the temperature of the environment respectively. Furthermore, $F_c$ is the thermal flux.

Under spherical symmetry, the numerical transient heat transfer problem is governed by:

$$\rho(r)c_p(r)\frac{\partial T(r,t)}{\partial t} = \frac{1}{r^2}\frac{\partial}{\partial r}\left(k(r)r^2\frac{\partial T(r,t)}{\partial r}\right) + \frac{\bar{Q}(r,t,\tau_0)}{V}, \qquad r \in [0, R_{dom}], \qquad t > 0 \quad (3)$$

where $R_{dom}$ and $V$ are the domain radius and PNP's volume. $\rho(r)$, $c_p(r)$, and $k(r)$ account for the radial dependence of the thermophysical properties across the different material domains, and $\bar{Q}(r,t,\tau_0)$ is the heat power defined in equation (1). The system is initially at thermal equilibrium,

$$T(r,0) = T_0(r) = T_\infty \quad (4)$$

and spherical symmetry imposes a zero-heat flux ($F_c$) at the centre,

$$\left.\frac{\partial T}{\partial r}\right|_{r=0} = 0 \quad (5)$$

At gold/water interface ($r = a$), a finite ITC ($G$) is introduced through the corresponding temperature-jump condition,

$$-k_{Au}\left.\frac{\partial T_-}{\partial r}\right|_{r=a} = G(T_- - T_+) = -k_{water}\left.\frac{\partial T_+}{\partial r}\right|_{r=a} \quad (6)$$

where $k_{Au}$ and $k_{water}$ are the thermal conductivities of gold and water respectively. At the outer boundary of the computational domain ($R_{dom}$), a Robin condition was imposed with a heat transfer coefficient $h = k_{water}/R_{dom}$. The numerical implementation additionally supports Dirichlet and Newmann conditions:

$$-k_{water}\left.\frac{\partial T}{\partial r}\right|_{r=R_{dom}} = h_{ext}(T(R_{dom},t) - T_\infty), \qquad h_{ext} = \begin{cases} 0 & (Neumann) \\ +\infty & (Dirichlet) \\ h & (Robin) \end{cases} \quad (7)$$

The resulting time-dependent heat equation was solved using an in-house finite-difference code based on the Crank-Nicolson scheme [31] . The implementation is formulated for a spherical geometry and allows the thermophysical properties and ITCs to be independently specified for each material. Additionally, the software implements the “Mie Scat” package [32–34] . that solves the light-sphere interaction problem and computes the absorption cross section. Further details on the code development are provided in Section S2 of the Supplementary Information.

The numerical implementation was validated by comparison with the analytical solution available for a gold sphere embedded in water. Details on this benchmark and the corresponding convergence analysis are provided in Section S2 of the supplementary information. The software used in this work is publicly available through Zenodo [35] .

## 3.Thermal confinement analysis

A. Spatial confinement analysis.

To quantify the spatial confinement of the generated heat, we consider the temporal envelope of the temperature increase,

$$\mathcal{T}(r) = \max_t[\Delta T(r,t)] \quad (8)$$

which represents the maximum temperature increase reached at each spatial position during the heating and cooling process. As illustrated in Figure 2, the envelope is constructed from the family of transient radial temperature profiles and allows for the interpretation of thermal confinement. Since the calculations consider excitation timescales differing by several orders of magnitude, the peak input power and the maximum temperature increase may vary substantially even for a constant input optical energy. Thus, to isolate the spatial distribution of heat from these amplitude variations, the envelope in the surrounding medium is normalized according to

$$\bar{\mathcal{T}}(r) = \frac{\mathcal{T}(r)}{\mathcal{T}(a^+)}, \qquad r \geq a \quad (9)$$

where $a^+$ numerically denotes the first computational node outside the NP. This reference choice is particularly convenient when finite ITCs are considered. In the CW limit, for a fixed absorbed power, the temperature profile in the surrounding medium is independent of the ITC, whereas a finite ITC modifies the temperature discontinuity across the interface and, consequently, the PNP temperature[19] . Therefore, the external surface temperature provides a common reference for comparing the effect of the ITC on the thermal confinement.

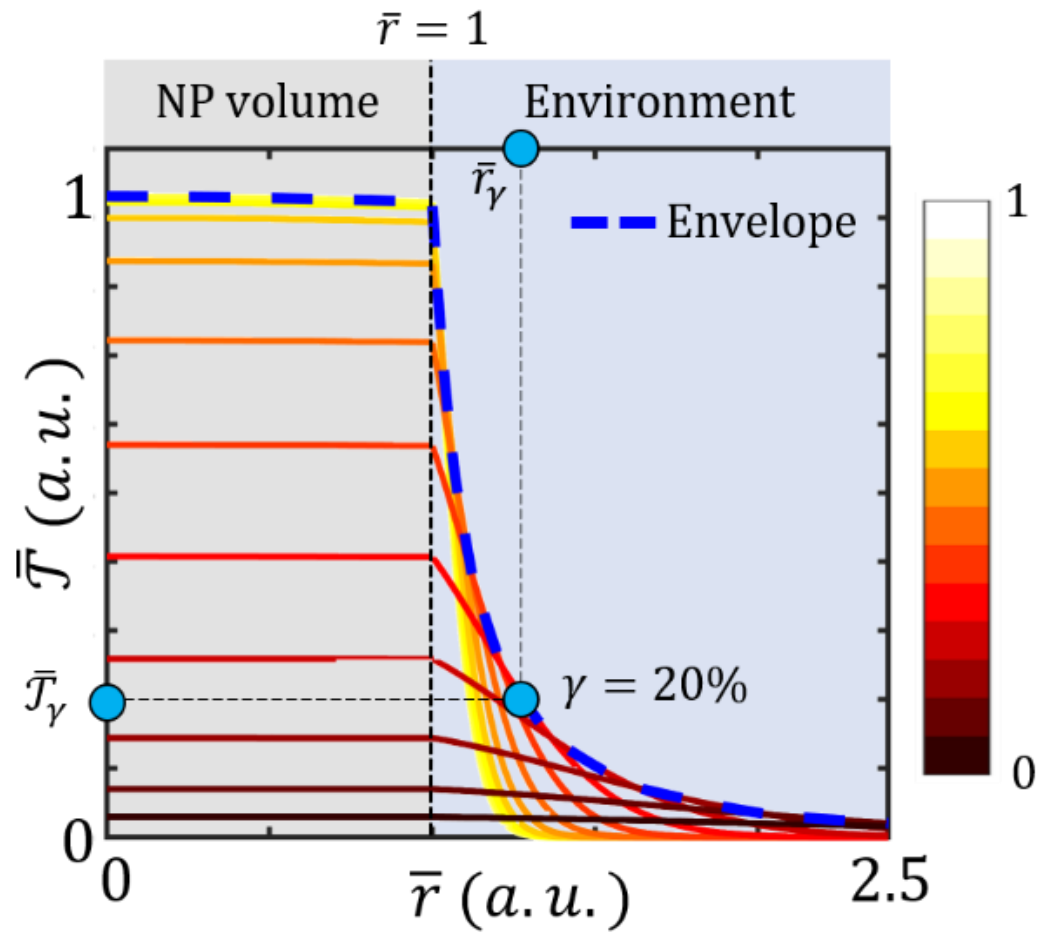

Figure 2. Representative example of the nanoparticle cooling process. The blue dashed line represents the envelope obtained as the temporal peak temperature for each spatial node. $\bar{r}_\gamma$ is the normalized radial distance for which the normalized temperature envelope has reached $\bar{\mathcal{T}}_\gamma$ i.e., a fraction $\gamma$ of its maximum value. The temperature envelope reaches its maximum at the particle surface, $\bar{r} = 1$.

To enable comparison between NPs of different sizes, the radial coordinate is additionally normalized to the particle radius, $\bar{r} = r/a$. From this normalization, the thermal confinement radius $\bar{r}_\gamma$ is defined as the normalized radial distance $\bar{r}$ at which the normalized envelope decreases to a fraction $\gamma$ of its value at the NP surface.

$$\bar{\mathcal{T}}(\bar{r}_\gamma) = \gamma, \qquad 0 < \gamma < 1 \tag{10}$$

As shown schematically in Figure 2, $\bar{r}_\gamma$ provides a direct measure of the spatial extent of the temperature increase into the surrounding medium. The representative case displayed in the figure corresponds to $\gamma = 20\%$. To quantify the effect of the ITC on the thermal confinement ($r_\gamma$), we define the relative confinement contrast as:

$$\epsilon_\gamma = \frac{\left|\bar{r}_\gamma^G - \bar{r}_\gamma^\infty\right|}{\bar{r}_\gamma^\infty} \cdot 100 \tag{11}$$

where the superscripts $G$ and $\infty$ stand for the parameters associated with $G = 200$ MW·K$^{-1}$·m$^{-2}$ [36] and $G = +\infty$ MW·K$^{-1}$·m$^{-2}$, respectively.

B. Transient analysis

To complement the spatial confinement analysis, we examine the transient thermal profiles, $\Delta T(\bar{r}, t)$, at specific radial distances $\bar{r}^{(j)} = 0, 1$ over the range of pulse durations considered. The temporal coordinate is normalized to the characteristic pulse duration, $\tau$, to enable direct comparison between excitation timescales as $\tilde{t} = t/\tau_0$. Furthermore, for each pulse duration, the transient temperature increase is normalized by its corresponding temporal maximum, i.e., the envelope evaluated at $\bar{r}^{(j)}$.

$$\overline{\Delta T}\left(\bar{r}^{(j)}, \bar{t}\right) = \frac{\Delta T\left(\bar{r}^{(j)}, \bar{t}\right)}{\mathcal{T}\left(\bar{r}^{(j)}\right)} \tag{12}$$

This normalization removes variations in the absolute temperature amplitude associated with the excitation timescale and allows the temporal shape of the thermal response to be compared consistently across the complete range of pulse durations.

Following a similar strategy than in the previous section, to analyse the impact that ITC has on the dynamic evolution of the thermal profiles, we define the dynamic thermal contrast as the relative difference between the temperature profiles obtained with finite and infinite ITC:

$$\epsilon_T\left(\bar{r}^{(j)}, \bar{t}\right) = \frac{\Delta T^G\left(\bar{r}^{(j)}, \bar{t}\right) - \Delta T^\infty\left(\bar{r}^{(j)}, \bar{t}\right)}{\mathcal{T}^\infty\left(\bar{r}^{(j)}\right)} \cdot 100 \tag{13}$$

and quantifies both the magnitude and the sign of the discrepancy, where the superscripts $G$ and $\infty$ denote finite and infinite ITC values. Under CW illumination, this equation reduces to the well-known steady-state thermal contrast [29] :

$$\epsilon_T = \frac{k}{aG} \cdot 100, \qquad r \leq a \tag{14}$$

## RESULTS

### 1. Thermal confinement in spherical NPs for generalized pulsed illumination.

Having defined the confinement radius $\bar{r}_\gamma$, we first investigate its dependence on PNP size and pulse duration. Figure 3a-f show the mapping of the confinement radius $\bar{r}_\gamma$ as a function of the particle radius $a$ and the log-transformed pulse duration $\tau_0$ for $\gamma = 0.1, 0.2, 0.3, 0.37, 0.4, 0.5$. The value $\gamma \approx 0.37$ is included as it corresponds to the conventional reference used for exponential decay. Black dashed lines indicate representative iso-$\bar{r}_\gamma$ contours.

A common behaviour is observed for all values of $\gamma$. Short excitation pulses produce strongly confined temperature distributions, whereas increasing the pulse duration progressively extends the heated region into the surrounding medium. This behaviour reflects the competition between the excitation timescale and thermal diffusion: when the pulse duration is short compared with the characteristic diffusion time, only a limited amount of heat propagates away from the nanoparticle during excitation. As the pulse duration increases, thermal diffusion becomes progressively more effective and $r_\gamma$increases.

For sufficiently long pulses, the confinement radius approaches the CW limit. In steady state, the normalized envelope of a spherical heat source follows $\bar{\mathcal{T}}(\bar{r}) = 1/\bar{r}$, and therefore it translates into $\bar{r}_\gamma = 1/\gamma$. Accordingly, the asymptotic values in Figures 3a–f are approximately $10, 5, 3.33, 2.70, 2.50$, and $2.00$ for $\gamma = 10\%, 20\%, 30\%, 37\%, 40\%$, and $50\%$, respectively, in excellent agreement with the analytical CW prediction.

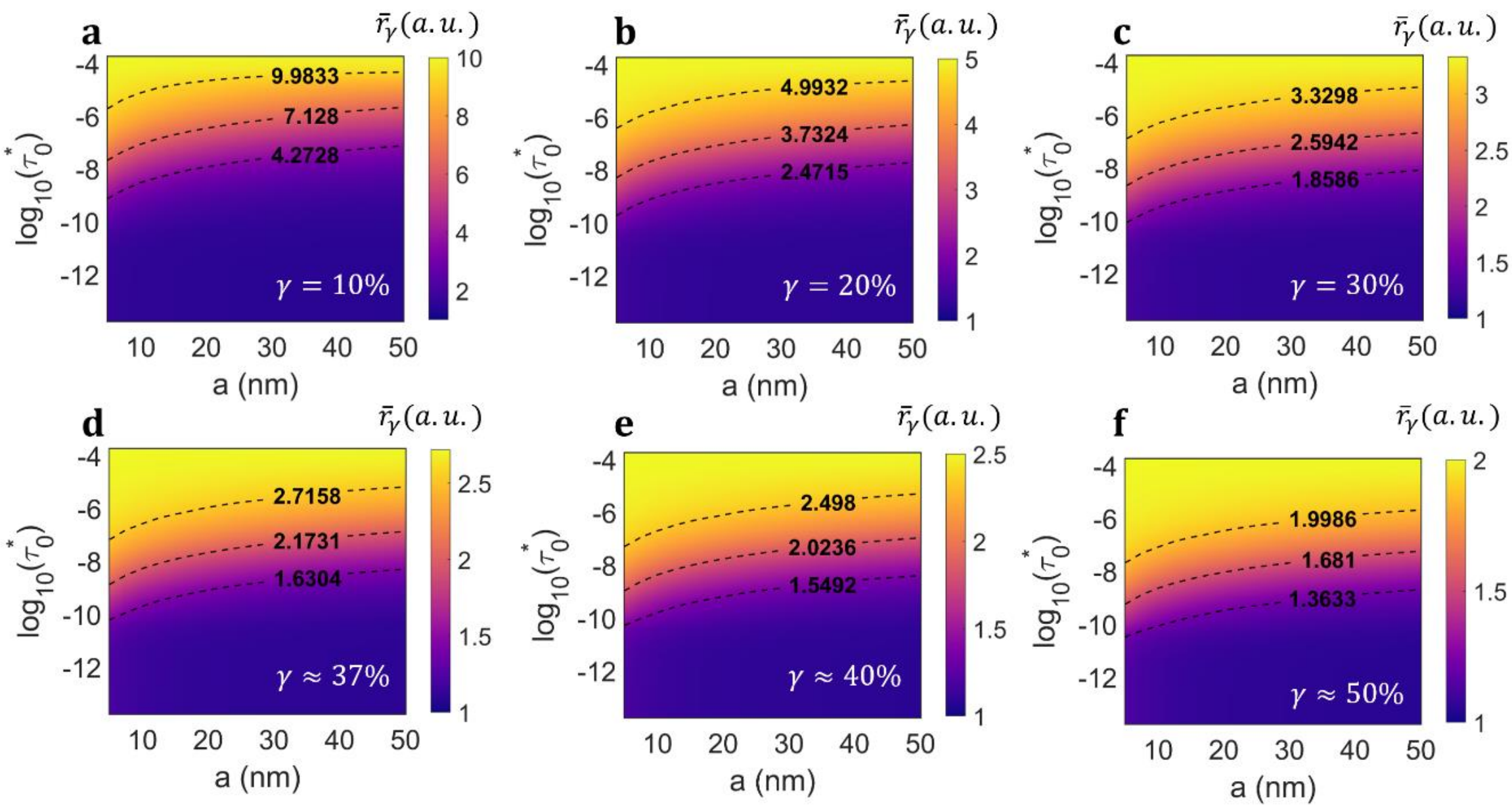

Figure 3. Study of the thermal confinement in a nanoparticle excited under generalized pulsed illumination. a)-f) Mapping of the thermal confinement distance, $\bar{r}_\gamma$, as a function of particle radius, $a$, and pulse duration in logarithmic scale $\log_{10} \tau_0^*$, for relevant threshold values $\gamma = 10, 20, 30, 37, 40, 50$ %. The black dashed lines show three representative iso-$\bar{r}_\gamma$ contours as a guide for the eye.

Furthermore, at a fixed pulse duration, the normalized confinement radius exhibits a weaker decrease with increasing particle size. Rather than originating from particle size independently, this trend reveals a universal scaling behaviour governed by the characteristic thermal diffusion time,

$$\tau_d = \frac{a^2}{\alpha_m},$$

where $\alpha_m$ is the thermal diffusivity of the surrounding medium. By reducing the parameter space to the dimensionless excitation timescale $\tau_0/\tau_d$, the confinement contours become robustly independent of the particle radius. In other words, increasing $a$ increases $\tau_d$ and therefore decreases $\tau_0/\tau_d$, shifting the system towards a more impulsive thermal regime and reducing the spatial extent of heating relative to the particle radius (see Figure S5 in Section 3 of the supplementary document).

Under this rescaling, the dependence on particle radius is strongly reduced and the iso-$\bar{r}_\gamma$ contours become approximately independent of $a$. This behaviour reduces substantially the computational effort required to explore the parameter space, since the thermal confinement can be characterized for a reference PNP size and then generalized to other radii through the appropriate time rescaling. In addition, the observed scaling indicates that the degree of thermal confinement can be effectively tuned through the duration of the incident light pulse.

**2. Effect of considering ITC on thermal confinement.**

Based on the scaling behaviour identified above, we now focus on a particle of a fixed radius $a = 25$ nm to isolate the influence of the ITC on thermal confinement. Figure 4a shows the confinement radius $\bar{r}_\gamma$ as a function of the envelope threshold $\gamma$ and pulse duration for perfect thermal contact, $G = +\infty$ mWK$^{-1}$m$^{-2}$. The iso-$\bar{r}_\gamma$ contours reveal three characteristic regimes denoted as R.I-R.III: the Quasi-Stationary Regime (R.I) for long-pulses, the Impulsive Regime (R.III) for ultrashort excitation and the transitional Confinement Regime (R.II) between both limits.

In R.I, $r_\gamma$ presents a very weak dependence on the pulse duration and approaches the CW behaviour discussed above. In contrast, in R.III, the confinement radius reaches its minimum values and becomes only weakly dependent on the pulse duration, reflecting the strongly confined character of ultrashort-pulse heating. Thus, most of the variation in $r_\gamma$ occurs within R.II, which defines the transition between diffusion-dominated and impulsive heating. This behaviour further confirms that the pulse duration can be an effective tool to tune the confinement of the generated thermal field.

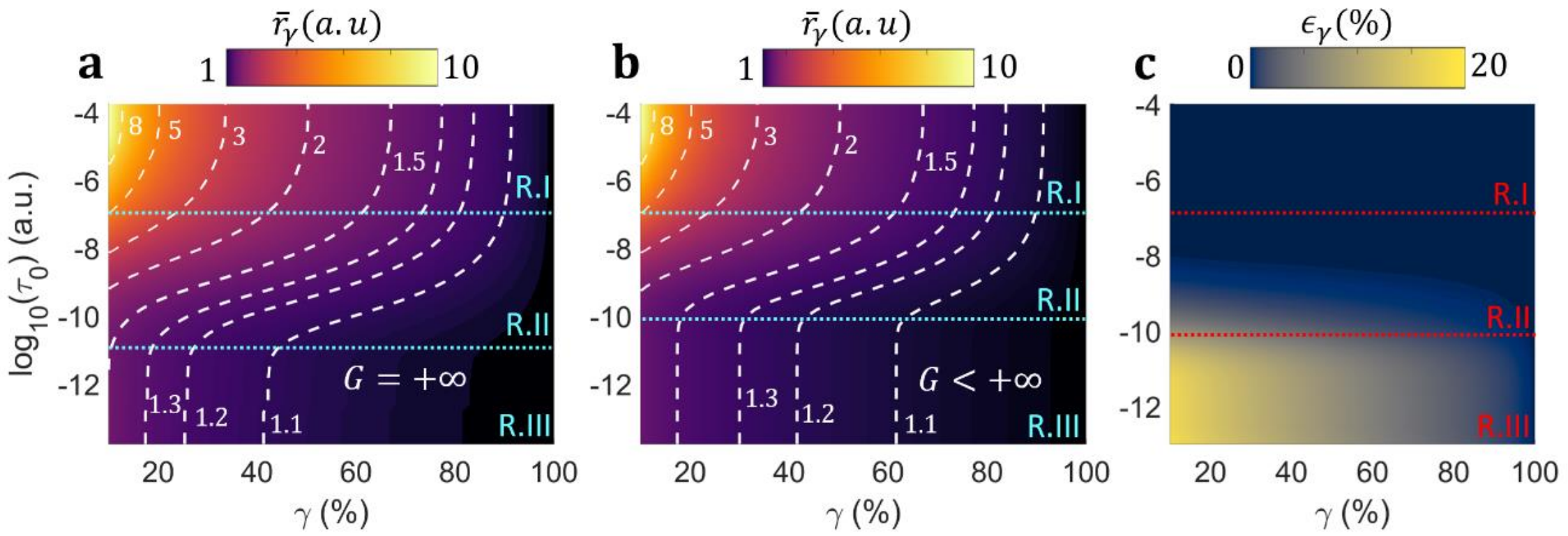

Figure 4. Mapping of the thermal confinement radius, $r_\gamma$, as a function of $\gamma$, and pulse duration in logarithmic scale $\log_{10}\tau_0$ for $G = +\infty$ MWK$^{-1}$m$^{-2}$ (a) and $G = 200$ MWK$^{-1}$m$^{-2}$ [36] (b). The white dashed lines show several reference contours as a guide for the eye. (c) Relative contrast $\epsilon_\gamma$ (equation 11) between the results in (a) and (b) as a function of the same parameters. R.I, R.II and R.III stand for first second and third confinement regions.

Figure 4b shows the $\bar{r}_\gamma$ map for a finite Au–water ITC, $G = 200$ MW K$^{-1}$m$^{-2}$ [36] . As it can be seen, R.I remains essentially unchanged, consistent with the fact that the normalized external temperature distribution is independent of $G$ for the CW limit. Differences progressively emerge as the pulse duration decreases, producing a modification of R.II. This transition region becomes narrower because of a systematic displacement of the iso-$r_\gamma$ contours in R.III. Therefore, for a given value of $\gamma$, finite ITC results in a larger confinement radius, indicating a more spatially extended normalized temperature envelope in the surrounding medium.

This effect can be understood from the additional timescale introduced by interfacial heat transfer. Under ultrashort excitation, a finite ITC delays the release of the energy stored in the PNP into the surrounding medium. The interface therefore effectively broadens the thermal excitation experienced by the environment, allowing heat to diffuse over a larger relative distance and consequently increasing $r_\gamma$. In the long-pulse limit, this interfacial timescale becomes negligible compared with the excitation timescale, and the finite- and infinite-ITC cases converge.

The magnitude of this effect is quantified in Figure 4c through the relative confinement contrast $\varepsilon_\gamma$ defined in equation 11. As expected, $\varepsilon_\gamma \to 0$ throughout R.I, whereas appreciable differences emerge across R.II and reach their largest value in R.III, with deviations going up to approximately 20%. Besides, the ITC-induced variation increases towards lower values of $\gamma$, corresponding to regions farther from the PNP surface. Thus, although the ITC has no relevant influence on the normalized spatial profile in the quasi-CW limit, it can significantly modify the thermal confinement predicted under ultrashort-pulse excitation.

Once the influence of the ITC on the spatial confinement of the thermal field is studied, we explore how this effect emerges from the thermal dynamics. For this purpose, the temperature evolution is analysed at two representative positions: immediately outside the nanoparticle surface, $\bar{r} = 1^{+}$, and at the nanoparticle centre, $\bar{r} = 0$.

Figures 5a and 5b show the normalized transient temperature increase at $\bar{r} = 1^{+}$ for infinite and finite ITCs, respectively. The maps recover the same three dynamical regimes identified from the confinement analysis. In the long-pulse regime, the temporal responses become nearly indistinguishable, whereas progressively larger changes in the shape of the thermal response emerge as the excitation enters the ultrashort regime. A finite ITC broadens the thermal response experienced by the surrounding medium, consistent with the larger confinement radii observed in Figure 4.

The corresponding dynamic thermal contrast at $\bar{r} = 1^{+}$ is shown in Figure 5c. In the long-pulse limit, $\varepsilon_T$ vanishes, recovering the expected CW behaviour in which the external temperature profile is independent of the ITC. As the pulse duration decreases, $\varepsilon_T$ becomes increasingly negative, reaching deviations larger than $60\%$ in the ultrashort regime. The negative sign indicates that a finite ITC reduces the instantaneous temperature increase in the surrounding medium relative to the perfect-contact case. This behaviour can be explained in terms of the ITC, which limits the rate at which the energy stored in the PNP is transferred to the environment.

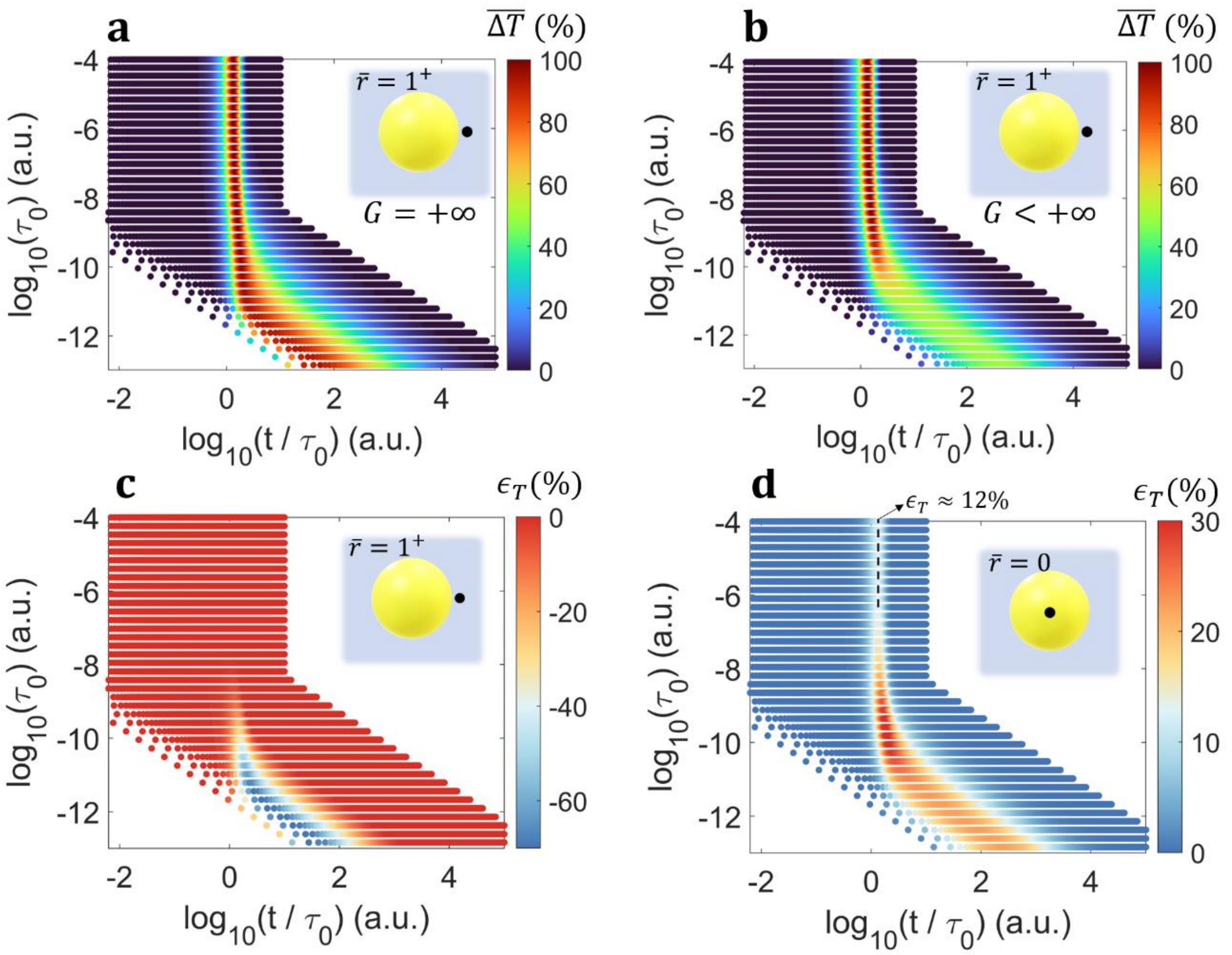

Figure 5. Normalized temperature profile dynamics for different pulse durations, comparing cases with infinite and finite ITC. Panels (a) and (b) show normalized temperature increase in the surrounding medium, just outside the nanoparticle surface ($\bar{r} = 1^+$), for the cases of (a) infinite ITC and (b) finite ITC. Panels (c) and (d) display the corresponding relative difference between the two cases, quantified by $\epsilon_T(t)$ for (c) the medium just outside the nanoparticle ($\bar{r} = 1^+$) and (d) the nanoparticle centre ($r = 0$).

The opposite behaviour is observed inside the nanoparticle. Figure 5d shows $\varepsilon_T$ evaluated at $\bar{r} = 0$, which remains positive throughout the explored parameter space. Thus, while finite ITC suppresses the transient temperature increase in the surrounding medium, it simultaneously increases the nanoparticle temperature. The contrast becomes particularly pronounced for ultrashort excitation, reaching values of approximately $30\%$. The corresponding normalized transient temperature increase at $\bar{r} = 0$ is shown in Figure S6.

Both observations originate from the same mechanism: a finite interfacial conductance hinders heat transfer across the nanoparticle–medium interface, causing energy to remain temporarily stored within the nanoparticle while reducing and delaying its release into the surrounding medium.

As discussed in the methods section, in the CW limit, the internal thermal contrast reduces to the analytical steady-state expression in equation 14. Using the parameters considered here, this expression yields $\epsilon_T \simeq 12\%$, in excellent agreement with the long-pulse plateau recovered in Figure 5d. This agreement establishes a direct connection between the transient description and the expected stationary limit.

**3. Feasibility study on thermal confinement control in ligand-driven applications.**

In this section, we build from the previous results obtained under the assumption of a constant ITC, a parameter that has been shown to play a relevant role in determining the degree of thermal confinement. To illustrate its practical relevance, we consider ligands commonly used for targeting tumour cells, whose characteristic lengths are well established. In particular, we focus on aptamers and antibodies, as they represent two clearly distinct size regimes [20,25] : aptamers typically span lengths from approximately 1 nm to 7 nm, whereas antibodies range from about 10 nm to 15 nm. For consistency, we fix the particle radius to $a = 25$ nm, noting that the same analysis can be readily extended to other particle sizes without loss of generality.

Figure 6 reproduces the results from Figure 4b, now expressed in terms of two independent variables: the pulse duration (on a logarithmic scale) and the confinement radius. The dependent variable, represented through a colour map, is the parameter $\gamma$. This representation provides a convenient way to identify the compatibility between ligand lengths and the thermal confinement landscape, with the goal of highlighting the accessible dynamic range for each ligand type.

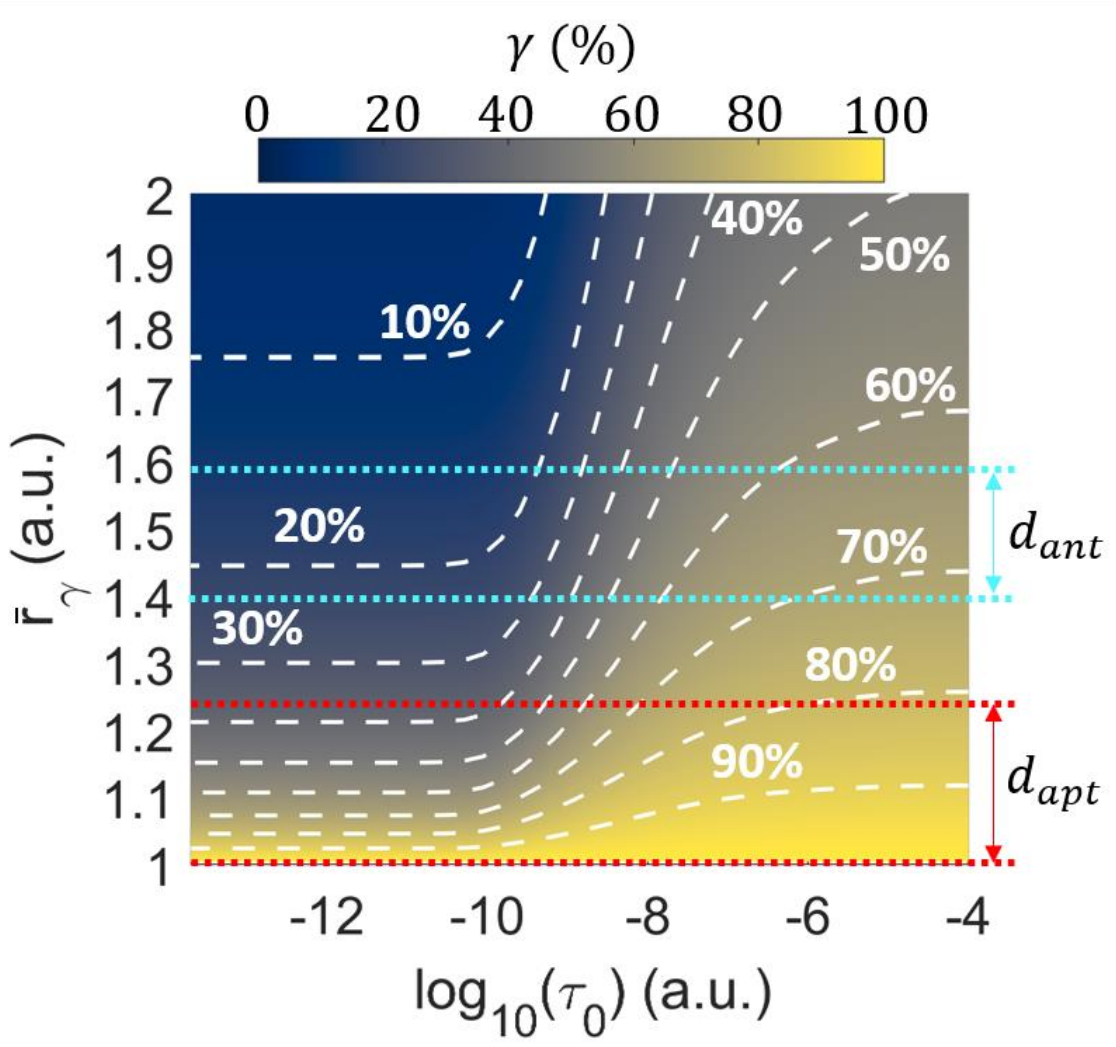

Figure 6. Map of the envelope threshold $\gamma$ as a function of the normalized thermal confinement radius $\bar{r}_\gamma$ and the logarithm of the pulse duration, $\log_{10}(\tau_0)$. White dashed lines represent iso-$\gamma$ contours ranging from 10% to 90%. Horizontal dotted lines indicate selected confinement radii reference regions for antibodies ($d_{ant}$) and aptamers ($d_{apt}$), respectively.

By normalizing ligand lengths with respect to the particle radius, we obtain the intervals $d_{apt} = [1.04, 1.24]$ for aptamers and $d_{ant} = [1.40, 1.60]$ for antibodies. These ranges are indicated in the figure by dashed red and light blue lines, respectively. As observed, the aptamer regime corresponds to regions closer to the particle surface, extending up to approximately 25% farther from it compared to the minimum distance. Within this interval, the isolines indicate that $\gamma$ varies from nearly 100% down to about 40%, reflecting a strong to moderate degree of thermal confinement.

In contrast, antibodies probe regions located further away from the particle surface, roughly between 40% and 60% of the normalized radial distance. In this case, the corresponding $\gamma$ values span a significantly broader and lower range, from around 70% down to approximately 20%. This indicates a substantially reduced thermal confinement, with heat spreading more efficiently into the surrounding medium.

Taken together, these results highlight a clear distinction between the two ligand types: aptamers operate within a regime where thermal confinement remains relatively strong over their entire spatial extent, whereas antibodies sample regions where confinement is progressively lost. Consequently, by tuning the pulse duration, it is possible to modulate how far the thermal field extends in space and time, effectively controlling whether the heating remains localized within the ligand region or extends beyond it. This provides a direct handle to tailor thermal delivery depending on the specific targeting strategy.

## CONCLUSIONS

In this work, we developed a numerical framework to quantify and interpret thermal confinement in spherical plasmonic nanoparticles under illumination conditions going from ultrashort pulses to the CW regime. By introducing the thermal confinement radius, $r_\gamma$, defined from the temperature-envelope distribution in the surrounding medium, the thermal confinement can be characterized independently of the input power, particle radius, and then absolute temperature increase. This approach provides a common metric for comparing different excitation timescales, nanoparticle sizes, and interfacial thermal conditions.

The results presented here indicate that pulse duration constitutes a relevant control parameter for thermal confinement. Ultrashort excitation produces highly localized heating around the nanoparticle, whereas increasing the pulse duration progressively extends the thermal field into the surrounding medium until the CW limit is recovered. The dependence on particle size can be accounted for by rescaling the characteristic thermal diffusion time, reducing the computational complexity of the analysis.

Furthermore, ITC introduces an additional timescale that becomes particularly relevant under ultrafast excitation. As expected, the consideration of a finite ITC significantly modifies both the spatial confinement and the transient thermal response at short pulse durations. In particular, ITC delays energy transfer from the nanoparticle to the surrounding medium, reducing the instantaneous external temperature while temporarily increasing the nanoparticle temperature.

From an application perspective, these results indicate that thermal confinement can be tailored not only through nanoparticle geometry and material properties, but also through pulse duration. The analysis of representative targeting ligands further illustrates how the relevant thermal length scale can be compared with the spatial dimensions of functional molecular layers, providing a simple criterion to assess whether heating remains localized within the ligand region or propagates into the surrounding environment.

More refined descriptions could incorporate the heterogeneous nature of interfacial thermal transport introduced by surface functionalization. Recent atomistic studies have shown that ligand chemistry and local solvent chemical properties can generate strongly nonuniform heat-transfer pathways across functionalized gold interfaces [26,27] . Combining spatially dependent or effective ligand-specific ITCs into the present framework could provide a more realistic and precise description of thermal confinement in biofunctionalized nanoparticles.

Ultimately, this theoretical framework provides a predictive tool that will contribute to the rational design of next-generation targeted photothermal therapies, enabling the precise spatiotemporal localization of heat to maximize therapeutic efficacy while reducing collateral damage in biological environments.

## AUTHOR CONTRIBUTIONS

Javier González-Colsa: Conceptualization, Methodology, Software, Formal analysis, Investigation, Data curation, Visualization, Writing – original draft, Writing – review & editing.

Dolores Ortiz: Methodology, Validation, Writing – review & editing.

José María Saiz: Methodology, Validation, Writing – review & editing.

Alfredo Franco: Writing – review & editing.

Fernando Bresme: Conceptualization, Methodology, Writing – review & editing, Supervision.

Francisco González: Project administration, Funding acquisition.

Pablo Albella: Project administration, Methodology, Writing – review & editing, Funding acquisition, Resources.

## DECLARATION OF COMPETING INTERESTS

The authors declare that they have no known competing financial interests or personal relationships that could have appeared to influence the work reported in this paper.

## ACKNOWLEDGEMENTS

P.A., F.G., D.O., J.M.S. and A.F. acknowledge support from the Spanish Ministerio de Ciencia e Innovación under project MOPHOSYS (grant no. PID2022-139560NB-I00). J. G.-C. acknowledges financial support from the Xunta de Galicia through the Programa de axudas de apoio á etapa de formación posdoutoral (ED481B-2026-093), funded by the Consellería de Educación, Ciencia, Universidades e Formación Profesional. F.B. thanks the UK Leverhulme Trust (Grant No. RPG-2018- 384).

## DATA AVAILABILITY

Numerical datasets supporting Figures 3-6 are available at Zenodo (https://doi.org/10.5281/zenodo.23372995).

The scientific software used to perform the numerical simulations is publicly available through Zenodo (https://doi.org/10.5281/zenodo.23084142).

# Supplementary information: Thermal Confinement in Plasmonic Nanoparticles under Generalised Pulsed Illumination

Javier González-Colsa[a,b,*], Dolores Ortiz[b], José María Saiz[b], Alfredo Franco[b], Francisco González[b], Fernando Bresme[c], Pablo Albella[b]

[a]CINBIO, University of Vigo, Campus Universitario de Vigo, Lagoas Marcosede, Vigo 36310, Spain.

*E-mail: javier.gonzalez.colsa@uvigo.gal.

[b]Group of Optics, Department of Applied Physics, University of Cantabria, 39005 Santander, Spain.

[c]Department of Chemistry, Molecular Sciences Research Hub, Imperial College London, London W12 0BZ, UK.

SECTION 1: Heat power density analysis

a) Functional Shape of the Heat Power Density

The process by which optical energy is transformed into heat by electronic absorption in metallic nanoparticles strongly depends on the pulse characteristics and the nanoparticle properties such as shape, size and materials. When it comes to the light-matter interaction problem, typically, the absorption cross section carries a major part of those properties. This is more adequate for sufficiently small nanoparticles compared to the excitation wavelength (quasistatic approximation), where the absorption becomes a simple closed expression [1] :

$$\sigma_{abs} = 4\pi k_m a^3 \mathbb{Im}\left[\frac{\epsilon_1 - \epsilon_2}{\epsilon_1 + 2\epsilon_2}\right] \tag{1.1}$$

where $k_m$ is the wavenumber in the surrounding medium. $a$ and $\epsilon$ are the nanoparticle radius and the permittivity respectively. The subscript 1 and 2 stand for nanoparticles and medium materials. The pulse properties, however, can be described as a temporal Gaussian profile as follows:

$$p(t) = \sigma_{abs} F x(t) = \frac{\sigma_{abs} F}{\tau_0} e^{-\pi\frac{t^2}{\tau_0^2}} \tag{1.2}$$

where $x(t) = \frac{1}{\tau_0} e^{-\pi\frac{t^2}{\tau_0^2}}$ is the normalized input temporal profile defined by the pulse length $\tau_0$. The factor $F$ represents the fluence, necessary to define the total pulse energy, $E$:

$$E = \int_{-\infty}^{+\infty} p(t)dt = \sigma_{abs} F \tag{1.3}$$

Even though electromagnetic absorption can be considered instantaneous, the gained electronic energy takes a finite time to be transferred to the nanoparticle lattice and thus it becomes determinant for sufficiently short pulses. This energy relaxation process is usually modelled as:

$$h(t) = \frac{1}{\tau_e} e^{-\frac{t}{\tau_e}} \Theta(t) \tag{1.4}$$

where $\Theta(t)$ is the Heaviside's function and $\tau_e$, the electronic relaxation time with typical value, $\tau_e = 1.7$ ps.

This formulation enables the interpretation of the light-to-heat conversion as a convolutional process, in which the driving signal is gaussian and the impulse response follows an exponential decay [2]. In essence, the total absorbed power density can be described as a convolution in time as follows:

$$Q(t, \tau_0) = p(t, \tau_0) * h(t) \tag{1.5}$$

Equation (1.5) can be solved into an integral form as a function of the error function ($\mathrm{erf}(x)$):

$$Q(t, \tau_0) = \sigma_{abs} F \frac{e^{\frac{\tau_0^2}{4\pi\tau_e^2}}}{2} \left[\frac{e^{-\frac{t}{\tau_e}}}{\tau_e}\right] \left[1 - \mathrm{erf}\left(\sqrt{\pi}\left(\frac{1}{2\pi}\frac{\tau_0}{\tau_e} - \frac{t}{\tau_0}\right)\right)\right] \tag{1.6}$$

Although the integral function in equation (1.6) is given in terms of the error function, the following form in terms of the complementary error function ($\mathrm{erfc}(x)$) will be used to ensure numerical stability:

$$Q(t,\tau_0) = \sigma_{abs} F \frac{e^{\frac{\tau_0^2}{4\pi\tau_e^2}}}{2}\left[\frac{e^{-\frac{t}{\tau_e}}}{\tau_e}\right]\mathrm{erfc}\left(\sqrt{\pi}\left(\frac{1}{2\pi}\frac{\tau_0}{\tau_e}-\frac{t}{\tau_0}\right)\right) \tag{1.7}$$

b) Asymptotic behaviour: a scale problem.

In numerical and analytical treatments involving the error function ($\mathrm{erf}(x)$), a critical issue arises due to its nonlinear saturation behaviour and the vast difference in the scales at which it transitions between its asymptotic limits. Specifically, the function exhibits an S-shaped (sigmoidal) profile, with extremely rapid changes near $x = 0$, but saturating very slowly as $x \to \pm\infty$, approaching $\pm 1$ asymptotically. This creates a scaling mismatch when modelling physical processes that span multiple orders of magnitude in space, time, or energy. In particular, the heat transfer process we study here is one of those cases, since $\tau_e = 1.7$ ps is fixed for gold. Therefore, it is reasonable to consider this unit for the central scale, the pulse length will range between the limit in which the particle can be considered a homogeneous heat source and the continuous range regime. Consequently, one can safely consider $\tau_0 \in (0.1\tau_e, +\infty)$, being aware of the energy normalization in each case since the energy must be sufficient to produce relevant particle heating. This reasoning supports the effort of looking for asymptotic approximations since when the pulse length is hugely deviated from the impulse relaxation time, equation (1.7) becomes impractical.

In this section, the asymptotic expressions for equation (1.7) will be derived in the cases of $\tau_0 \gg \tau_e$ and $\tau_0 \ll \tau_e$, that for gold corresponds to ns or larger pulses and femtosecond illumination respectively. We will demonstrate the asymptotic expressions from the definition of convolution and from equation (1.7) to show its consistency.

- Case $\tau_0 \ll \tau_e$:

  -Derivation from convolution:

  Let us consider the normalized convolutional expression (without the total energy term) in equation (1.5). Therefore, we can write

$$\bar{Q}(t) = \frac{1}{\tau_0\tau_e}\int_{-\infty}^{+\infty} e^{-\frac{\tau}{\tau_e}}\,\Theta(t)\, e^{-\pi\frac{(t-\tau)^2}{\tau_0^2}}\, d\tau \tag{1.8}$$

  Since $\tau_0 \ll \tau_e$, $e^{-\pi\frac{(t-\tau)^2}{\tau_0^2}}\Big/\tau_0 \Rightarrow \delta(t-\tau) = \delta(\tau - t)$. In this approximation, equation (1.8) becomes:

$$\bar{Q}(t) = \frac{1}{\tau_e}\int_{-\infty}^{+\infty} e^{-\frac{\tau}{\tau_e}}\,\Theta(t)\delta(\tau - t)d\tau = \frac{1}{\tau_e}e^{-\frac{t}{\tau_e}}\Theta(t) = h(t) \tag{1.9}$$

-Derivation from expression (1.7):

Given the scaling contrast between $\tau_0$ and $\tau_e$, we can write $\tau_0 = \eta\tau_e$. where $\eta \ll 1$. Besides, for the same reason, it is reasonable rescaling equation (1.6) by using $t = \gamma\tau_e$ where $|\gamma| \sim 1$, i.e., $t$ share the same order of magnitude than $\tau_e$. With this rescaling, equation (1.7) can be simplified into:

$$Q(t,\tau_0) = \sigma_{abs} F \frac{e^{\frac{\eta^2}{4\pi}}}{2}\left[\frac{e^{-\gamma}}{\tau_e}\right] \operatorname{erfc}\left(\sqrt{\pi}\left(\frac{1}{2\pi}\eta - \frac{\gamma}{\eta}\right)\right) \tag{1.10}$$

Given the properties of $\eta$ and $\gamma$, it is clear that:

$$\lim_{\eta\to 0} e^{\frac{\eta^2}{4\pi}} = 1$$

$$\lim_{\eta\to 0} K(\eta,\gamma) = \lim_{\eta\to 0} \sqrt{\pi}\left(\frac{1}{2\pi}\eta - \frac{\gamma}{\eta}\right) = -\sqrt{\pi} \lim_{\eta\to 0}\left(\frac{\gamma}{\eta}\right) = \begin{cases} +\infty, & \text{if } \gamma < 0 \\ 0, & \text{if } \gamma = 0 \\ -\infty, & \text{if } \gamma > 0 \end{cases}$$

And thus:

$$\lim_{\eta\to 0} \operatorname{erfc}\big(K(\eta,\gamma)\big) = \begin{cases} 0, & \text{if } \gamma < 0 \\ 1, & \text{if } \gamma = 0 \\ 2, & \text{if } \gamma > 0 \end{cases}$$

This allows the rewrite equation (1.10) into:

$$Q(t,\tau_0) = \sigma_{abs} F \left[\frac{e^{-\frac{t}{\tau_e}}}{\tau_e}\right] \cdot \begin{cases} 0, & \text{if } t < 0 \\ \frac{1}{2}, & \text{if } t = 0 \\ 1, & \text{if } t > 0 \end{cases} \tag{1.11}$$

If we consider the definition of the Heaviside's function from the weak limit of regular smooth functions, in which $\Theta(0) = 1/2$, then equation (1.11) can be rewritten as:

$$Q(t,\tau_0) = \sigma_{abs} F \frac{e^{-\frac{t}{\tau_e}}}{\tau_e}\Theta(t) = \sigma_{abs} F\, h(t) \tag{1.12}$$

And, therefore, we recover the impulse response rescaled through the total energy.

- Case $\tau_0 \gg \tau_e$:

-Derivation from convolution:

Let us consider the normalized convolutional expression again (equation (1.8)). Taking advantage of the convolutional commutative property, we can write:

$$\bar{Q}(t) = \frac{1}{\tau_0\tau_e}\int_{-\infty}^{+\infty} e^{-\pi\frac{\tau^2}{\tau_0^2}} e^{-\frac{t-\tau}{\tau_e}}\Theta(t-\tau)d\tau \tag{1.13}$$

Since $\tau_0 \gg \tau_e$, $e^{-\frac{t-\tau}{\tau_e}}\Theta(t-\tau)/\tau_e \Rightarrow \delta(t-\tau) = \delta(\tau - t)$. In this approximation, equation (1.8) becomes:

$$\bar{Q}(t) = \frac{1}{\tau_e}\int_{-\infty}^{+\infty} e^{-\pi\frac{\tau^2}{\tau_0^2}}\,\delta(\tau - t)d\tau = \frac{1}{\tau_0}e^{-\pi\frac{\tau^2}{\tau_0^2}} = x(t) \tag{1.14}$$

-Derivation from expression (1.7):

As in the previous case, the scaling contrast between $\tau_0$ and $\tau_e$ justifies the rescaling of $\tau_0 = \eta\tau_e$. where $\eta \gg 1$. Besides, for the same reason, it is reasonable rescaling equation (1.7) by using $t = \gamma\tau_0$ where $|\gamma| \sim 1$, i.e., $t$ share the same order of magnitude than $\tau_0$. With this rescaling, equation (1.7) can be simplified into:

$$Q(t,\tau_0) = \sigma_{abs}F\frac{e^{\frac{\eta^2}{4\pi}}}{2}\left[\frac{e^{-\eta\gamma}}{\tau_e}\right]\mathrm{erfc}\left(\sqrt{\pi}\left(\frac{1}{2\pi}\eta - \gamma\right)\right) \tag{1.15}$$

Given the properties of $\eta$ and $\gamma$:

$$\lim_{\eta\to+\infty} K(\eta,\gamma) = \lim_{\eta\to+\infty}\sqrt{\pi}\left(\frac{1}{2\pi}\eta - \gamma\right) = \left(\frac{1}{2\sqrt{\pi}}\right)\lim_{\eta\to+\infty}\eta = +\infty$$

This limit justifies considering the following asymptotic form for the complementary error function:

$$\mathrm{erfc}(x) \sim \frac{e^{-x^2}}{\sqrt{\pi}x} = \frac{e^{-\pi\left(\frac{1}{2\pi}\eta-\gamma\right)^2}}{\pi\left(\frac{1}{2\pi}\eta - \gamma\right)} \approx 2\frac{e^{-\pi\left(\frac{1}{2\pi}\eta-\gamma\right)^2}}{\eta}$$

By combining the approximation into equation (1.15), it transforms into:

$$Q(t,\tau_0) = \frac{\sigma_{abs}F}{\tau_0}e^{-\pi\frac{t^2}{\tau_0^2}} = \sigma_{abs}Fx(t) \tag{1.16}$$

c) Validity range of asymptotic approximations

Although closed-form expressions for the limits of the convolution have been derived in the previous sections, these expressions rely on limiting procedures and asymptotic relations, valid under specific assumptions about the temporal relationship between the pulse and the exponential decay. Therefore, it is necessary to assess the accuracy of these approximations over a broad range of parameters to determine the extent of their validity and to establish clear criteria for selecting the appropriate expression in each regime of interest.

To quantitatively assess this, we compute expression (1.7) for a set of input pulse lengths $\tau_0$ keeping the impulse response length constant $\tau_e = 1.7$ ps (material fixed, gold). We, then, compare those functions against the asymptotic approximations in equations (1.9), (1.14). To determine the degree of similarity between the convolutional responses and the asymptotic expressions, the following metrics (relative mismatch) were used:

$$\delta_{\|\cdot\|_\infty} = \frac{\sup_t|\bar{Q}(t,\tau_0) - \phi(t)|}{|\phi(t_{max})\,|} \tag{1.17}$$

$$\delta_{\|\cdot\|_{L^1}} = \frac{\int_{-\infty}^{+\infty} |\bar{Q}(t,\tau_0) - \phi(t)| \, dt}{\int_{-\infty}^{+\infty} |\phi(t)| \, dt} \tag{1.18}$$

$$\delta_{\|\cdot\|_{L^2}} = \left( \frac{\int_{-\infty}^{+\infty} |\bar{Q}(t,\tau_0) - \phi(t)|^2 \, dt}{\int_{-\infty}^{+\infty} |\phi(t)|^2 \, dt} \right)^{\frac{1}{2}} \tag{1.19}$$

where $t_{max}$ is the value for which $\bar{Q}(t,\tau_0)$ reaches its maximum for a given $\tau_0$. Besides, the reference function $\phi(t)$ expresses:

$$\phi(t) = \begin{cases} h(t) \; if \; \tau_0 \ll \tau_e \\ x(t) \; if \; \tau_0 \gg \; \tau_e \end{cases}$$

For its part, equation (1.17) measures the maximum absolute value of the difference between the two functions over the entire domain. It identifies the instant where the deviation is the largest. Only the worst mismatch matters, regardless of the behaviour elsewhere. This norm is particularly useful when it is important to guarantee that the maximum error at any point in time remains controlled. On the other hand, equation (1.18) measures the total accumulated error between the two functions by integrating the absolute value of their difference across the domain. This accumulated error is normalized by the integral of the absolute value of the reference function, thus providing a relative measure of the mismatch. It is appropriate when the interest lies in minimizing the overall deviation, without emphasizing isolated large errors. Finally, equation (1.19) compares the two functions by integrating the square of their difference and then taking the square root. Since the errors are squared before integration, this metric penalizes larger deviations more heavily than smaller ones. In this formulation, the measure is normalized by the $\mathfrak{L}^2$ norm of phi, yielding a relative mismatch. We employed these metrics because they allow a progressive understanding of the importance of localized large errors. In particular, they enable the study of the mismatch between functions from a highly localized to a more global perspective. This approach makes it possible to assess the quality of the approximations not only with respect to the approximated function itself but also considering the structure of the limiting function. In this way, **one can interpret the adequacy of the approximation in relation to the complexity of the limit function**.

A detailed analysis of these behaviours is provided in Figure S1. As a general remark, the same rescaling procedure discussed in the previous section has been adopted, where time is expressed in units of $\tau_e$. This choice addresses the inherent scaling challenges posed by the highly disparate characteristic times involved in the problem. We first examine the regime $\tau_o \ll \tau_e$, focusing on the range $\tau_0 \in [1/60\,\tau_e, \tau_e]$. To better capture the scaling effects, an inverse normalization, $\tau_e/\tau_0$ has been employed, providing a more intuitive representation of the temporal discrepancies. As illustrated in Figure S1a, the convolutive response rapidly converges toward the form of the response function $h(t)$ as $\tau_e/\tau_0$ grows. However, it is important to emphasize that the limiting behaviour does not correspond to an ordinary function, but rather to a distribution featuring a first-order discontinuity at $t = 0$. This nuance becomes particularly evident upon closer inspection of Figure S1b.

When analysing the behaviour of the $l^\infty$-norm error, one observes that it does not significantly decrease with decreasing $\tau_0$, instead approaching a horizontal asymptote around 0.5 for arbitrarily small values of $\tau_0$. This phenomenon can be explained by the structural features of both the asymptotic distribution and the convolutive approximant: due to their profiles, the maximum relative discrepancy (as defined in Eq. (1.17)) consistently occurs near $x = 0$, and asymptotically tends to 1/2 as $\tau_0 \rightarrow 0$ . Consequently, these findings suggest that the $\delta_{\|\cdot\|_\infty}$ metric is not well-suited to fully characterize the relative discrepancy in this context, as it is overly sensitive to the behaviour at the discontinuity. Instead, it would be more appropriate to consider alternative metrics that account for the error distribution across the entire domain. When it comes to metrics $\delta_{\|\cdot\|_{L^1}}$ and $\delta_{\|\cdot\|_{L^2}}$in Figure S1b, it can be seen that $\delta_{\|\cdot\|_{L^2}} > \delta_{\|\cdot\|_{L^1}}$. This trend is naturally derived from expression (1.19), since it enhances the effect of the localized discrepancy at $t = 0$.

To establish a quantitative validity threshold for the errors, we will refer to the study presented in [2] . In that work, the use of the exponential approximation is justified for a characteristic pulse duration of approximately 200 fs, leading to a corresponding normalized characteristic time $\tau_e/\tau_0 = 8.5$. For this value in Figure S1b, we find $\left(\delta_{\|\cdot\|_{L^1}}, \delta_{\|\cdot\|_{L^2}}\right) \approx (6, 16)$ %. Based on those values, $\delta_{\|\cdot\|_{L^1}}$will be considered as a reference, since it informs about discrepancies across the entire domain with reduced localized discrepancies enhancement. Thus, we will assume a **maximum error threshold of 6% for this metric**.

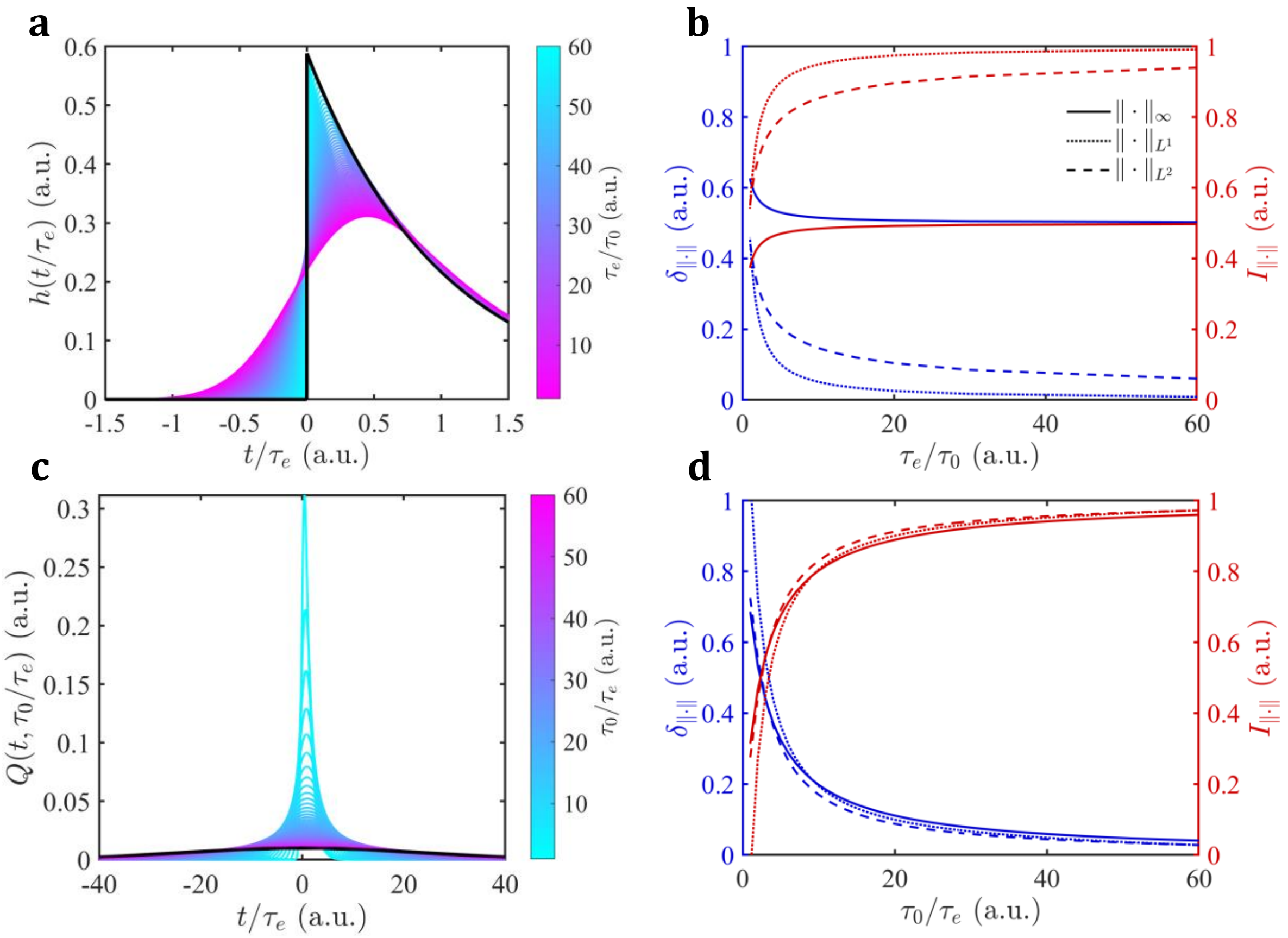

Figure S1. Asymptotic behaviour of the convolutive heat-source model in the short- and long-pulse limits. (a) Rescaled temporal profiles for increasing values of $\tau_e/\tau_0$, showing the progressive convergence of the convolutive response toward the electron–phonon response function $h(t/\tau_e)$(black line). (b) Corresponding normalized discrepancies evaluated using the $L_\infty$, $L_1$, and $L_2$norms as a function of $\tau_e/\tau_0$. (c) Analogous rescaled temporal profiles in the opposite limit, for increasing $\tau_0/\tau_e$, illustrating convergence toward the Gaussian excitation profile. (d) Corresponding $L_\infty$, $L_1$, and $L_2$discrepancy metrics as a function of $\tau_0/\tau_e$. In panels (a) and (c), the colour scale indicates the ratio between the characteristic pulse duration and the electron–phonon relaxation time.

Following the same criteria and focusing now on Figure S1c, a behaviour similar to the previous case can be observed, with the difference that the relative variable is now $\tau_0\ /\tau_e$. It can be seen that as this value increases, the convolutive response tends toward the limiting function. This is confirmed in Figure S1d, where all the metrics converge rapidly to zero as $\tau_0\ /\tau_e$ increases. This result is reasonable, given that, except near $t\ =\ 0$, no significant localized discrepancies are observed, and the limiting function remains smooth throughout the domain. Based on the previously established criterion for $\delta_{\|\cdot\|_{L^1}}$ we can consider that the convolutive approximation is adequate when $\tau_0/\tau_e \approx 33$.

As additional information, if we were to take the metrics $\delta_{\|\cdot\|_\infty}$ and $\delta_{\|\cdot\|_{L^2}}$ as references, the corresponding threshold values would be $\tau_0/\tau_e \approx 36$ and $\tau_0/\tau_e \approx 29$, respectively. Therefore, assuming $\delta_{\|\cdot\|_{L^1}}$ as the standard metric, we can conclude that the validity range for the convolutive response (Equation (1.7)) is given by $\tau_0\ \in\ [0.1\tau_e, 33\tau_e]$. As a conclusion, we can define the absorbed power as a piecewise function:

$$Q(t,\tau_0) = \begin{cases} \sigma_{abs}F\dfrac{e^{-\frac{t}{\tau_e}}}{\tau_e}\Theta(t) & \text{if } \tau_0 \in [\tau_{-\infty}, 0.1\tau_e) \\ \sigma_{abs}F\dfrac{e^{\frac{\tau_0^2}{4\pi\tau_e^2}}}{2}\left[\dfrac{e^{-\frac{t}{\tau_e}}}{\tau_e}\right]\operatorname{erfc}\left(\sqrt{\pi}\left(\dfrac{1}{2\pi}\dfrac{\tau_0}{\tau_e}-\dfrac{t}{\tau_0}\right)\right) & \text{if } \tau_0 \in [0.1\tau_e, 33\tau_e] \\ \dfrac{\sigma_{abs}F}{\tau_0}e^{-\pi\frac{t^2}{\tau_0^2}} & \text{if } \tau_0 \in (33\tau_e, \tau_{+\infty}] \end{cases} \tag{1.20}$$

where $\tau_{-\infty}$ and $\tau_{+\infty}$ minimum and maximum validity ranges for the exponential and Gaussian approximations, respectively. In the case of $\tau_{-\infty}$, a value for which the electron gas is thermalized can be assumed. However, this value depends on the environmental conditions and on the size of the nanostructure and therefore must be changed depending on the considered case [3] .

On the other hand, $\tau_{+\infty}$ can be considered as inifinite from a theoretical perspective, as the Gaussian temporal profile asymptotically approaches a constant intensity value upon broadening. However, it is noteworthy, that if the total absorbed energy remains constant, the function $Q(t,\tau_0) \to 0$. Therefore, the continuous wave (CW) condition represents a physically distinct situation from that described in Equation (1.20), where the Gaussian branch must be rescaled in energy as the pulse length increases. If we assume that fluence scales with illumination time as $F = I\tau_o$, beyond a certain $\tau_0$, where $I$ is the intensity, the absorbed power can be rewritten as:

$$Q(t,\tau_0)_+ = \frac{\sigma_{abs}F}{\tau_0}e^{-\pi\frac{t^2}{\tau_0^2}} = \sigma_{abs}Ie^{-\pi\frac{t^2}{\tau_0^2}}$$

Therefore, the limit:

$$\lim_{\tau_0\to+\infty} Q(t,\tau_0)_+ = \sigma_{abs}I\lim_{\tau_0\to+\infty} e^{-\pi\frac{t^2}{\tau_0^2}} = \sigma_{abs}I \tag{1.21}$$

, which fits with the absorbed power for a nanostructure illuminated by a continuous wave[4] .

SECTION 2: Matlab FD implementation for spherical metallic nanoparticles in a homogeneous medium.

The transient heat-transfer problem described in the main text was solved using an in-house finite-difference solver implemented in MATLAB. Under spherical symmetry, the computational domain is reduced to the radial coordinate, $0 \leq r \leq R$, and discretized into $N_r$ spatial nodes with spacing $\Delta r$. The thermophysical properties $\rho(r)$, $c_p(r)$, and $k(r)$ are assigned according to the material within each radial region.

Time integration is performed using the Crank–Nicolson scheme, with a temporal discretization $t_n = n\Delta t$. The volumetric heat-generation rate, $Q(r,t)$, is introduced as the source term and is restricted to the absorbing material domain.

Material interfaces are explicitly included in the radial discretization. At each interface, continuity of the heat flux is imposed, while a finite interfacial thermal conductance, ($G_i$), allows for temperature discontinuity across adjacent materials. In the limit ($G_i \to +\infty$), the temperature becomes continuous across the interface, recovering the perfect thermal-contact condition.

At the centre of the sphere, spherical symmetry imposes a zero radial temperature gradient. At the outer boundary $r = R_{dom}$, the solver supports Dirichlet, Neumann, and Robin boundary conditions. The specific boundary condition employed in each calculation is stated where appropriate.

The spatial domain, spatial and temporal discretizations, material parameters, source function, and boundary conditions can be independently specified. The complete implementation is publicly available at Zenodo [5] . The numerical accuracy of the solver and the choice of spatial and temporal discretizations are assessed below through convergence and validation tests.

Model validation

The model has been validated for the case of a gold sphere in water, which represents the paradigmatic scenario in thermoplasmonics. Continuous and pulsed illumination regimes have been investigated, including femtosecond (exponential) and nanosecond (Gaussian) excitations. The results were benchmarked against analytical solutions, specifically, in the case of a continuous-wave illumination and femtosecond pulsed excitation

$$T_{CW} = \begin{cases} \dfrac{Q}{6k_p}(a^2 - r^2) + \dfrac{Qa^2}{3k_m} + T_\infty & r \leq a \\ \dfrac{Qa^3}{3k_m r} + T_\infty & r > a \end{cases} \quad (2.1)$$

$$T_{CW}^{G} = \begin{cases} \dfrac{Q}{6k_p}(a^2 - r^2) + \dfrac{Qa^2}{3k_m} + \dfrac{Qa}{3G} + T_\infty & r \leq a \\ \dfrac{Qa^3}{3k_m r} + T_\infty & r > a \end{cases} \quad (2.2)$$

-Validation for CW illumination.

To ensure the accuracy of the code, the solution for a set of spheres is validated against the reference solution in equation (2.1) and (2.2). The numerical calculations are performed by discretizing the spherical heat conduction problem with explicit control of the spatial and temporal steps. For each simulation, the particle radius is varied as $a$ = {5, 20, 35, 50} nm. The number of radial nodes inside the particle is chosen as $N_p$ = {5, 10, 20, 40, 80, 160, 320}. The size of the computational domain is set as $R_{sim}$ = 50 · $a$, so that the far-field boundary condition is correctly imposed. The number of nodes in the surrounding medium is then computed according to $N_e = round((R_{sim} - a)/a \cdot N_p)$, which ensures that the spatial discretization inside and outside the particle are consistent. As a result, larger spheres require proportionally larger simulation regions and thus more external nodes. For the temporal discretization, we fix $N_t$ = 8000 steps and $t_{max}$ = 10 µs, so that the dynamics can be resolved over a sufficiently long-time window for all particle sizes. Once the numerical solution is obtained, it is benchmarked against the analytical steady-state solution for a sphere under continuous-wave illumination (Eq.2.1, 2.2). The discrepancy is quantified by means of the relative L2-error norm in Eq. 1.19, which measures the global error between numerical and analytical profiles. These conditions are applied for all boundary conditions.

+Dirichlet condition (isothermal).

For the Dirichlet condition, the temperature at the last node of the computational domain is forced to remain constant and equal to the ambient temperature, i.e., $T(R_{sim}, t) = T_{amb}$, for all times $t$. This approach guarantees that the far-field condition is imposed exactly at the numerical boundary, independently of the size of the simulation region. Although physically the surrounding medium extends to infinity, in practice the numerical domain is truncated at a finite radius $R_{sim}$. By fixing the temperature at this boundary, the solver emulates the presence of an infinite reservoir that acts as a heat sink.

Figure S2 shows the numerical performance of the method for different particle diameters ($D$ = 10, 40, 70, and 100 nm). Panels (a) and (b) quantify the relative error with respect to the analytical continuous-wave solution, using the $L^2$ norm of the full temperature profile $\delta(T(r))$ and the relative error in the maximum temperature $\delta(T_{max})$, respectively. Panel (c) reports the corresponding simulation times $t_{sim}$.

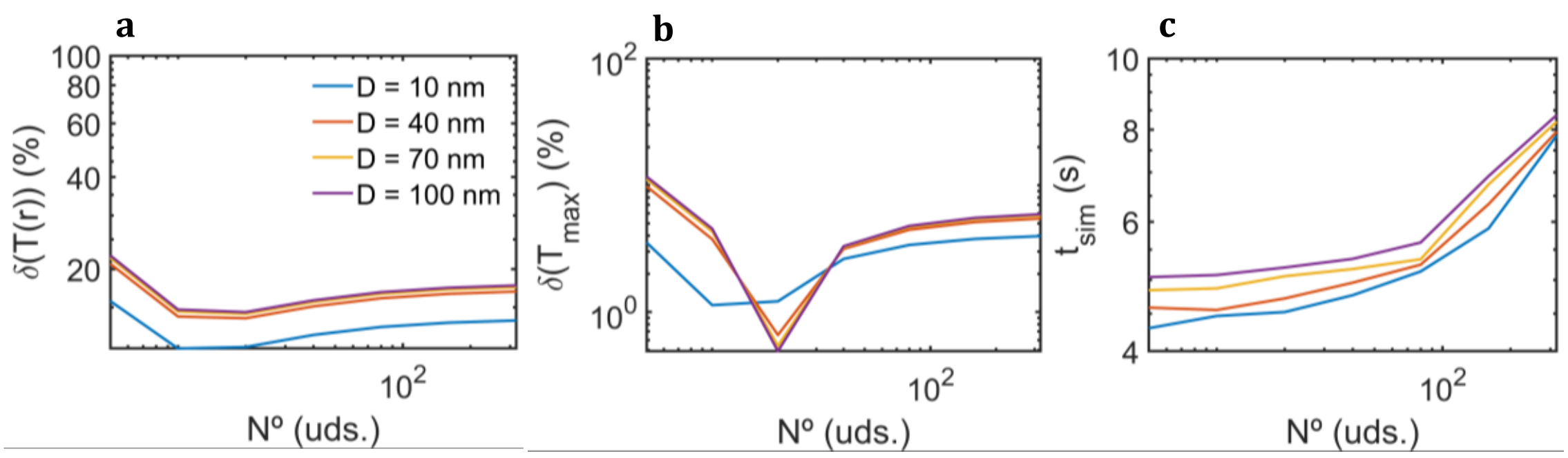

Figure S2. Relative error of the numerical solver compared to the analytical CW solution for different particle diameters ($D$ = 10, 40, 70, 100 nm). (a) Relative global error of the temperature profile $\delta(T(r))$, (b) error in the maximum temperature $\delta(T_{max})$, and (c) simulation time $t_{sim}$ as a function of the number of nodes inside the particle.

In both error metrics, a clear non-monotonic behaviour is observed: the error decreases as the number of nodes increases, reaches a minimum, and then increases again for the densest discretization. For example, in the case of D = 40-100 nm, the global error decreases from $\sim 20-30\%$ at $N_p = 5$ to $\sim 10\%$ at $N_p \approx 20$, before rising again towards $\sim 15-20\%$ for $N_p \geq 160$. A similar trend is found for $\delta(T_{max})$, where the error drops below 1% at intermediate discretizations ($N_p \approx 20-40$) but grows again towards ~5–10% at larger $N_p$.

This behaviour is directly related to the imposed Dirichlet boundary condition at the outer edge of the simulation domain: since the temperature is fixed at $T_{amb}$, there exists an optimal mesh density that balances the accurate representation of the spherical heat kernel with the truncation effects introduced by the finite boundary. When the mesh becomes too dense, the boundary condition dominates the discretization error, producing the observed increase. In addition, fixing the temperature at a finite distance systematically pulls down the entire numerical temperature profile compared to the ideal infinite-domain solution. The result is a global underestimation of the thermal field, with the discrepancy being more pronounced at larger radii, where the artificial constraint forces the solution to decay faster than it physically should.

Panel (c) shows that the simulation time grows approximately quadratically with the number of nodes, ranging from $\sim 4-5$ s at $N_p = 5$ to $\sim 8-9$ s at $N_p = 320$. The dependence on particle size is mild: for a fixed Np, the spread in simulation time between D = 10 nm and D = 100 nm is less than ~1 s. This weak dependence indicates that the computational cost is not driven by the size of the particle itself, but rather by the recursive solution of the linear systems. Since the matrices assembled for different diameters have the same dimensions (determined by $N_p$ and $N_e$), the observed size dependence is attributed to the conditioning and resolution process of the system rather than to the discretization.

Thus, the Dirichlet boundary condition can yield accurate results provided that the simulation domain is chosen sufficiently large to minimize the artificial influence of the fixed temperature at the outer boundary. In practice, this requires setting the external radius of the computational domain as a multiple of the particle radius, ensuring that the imposed ambient temperature does not distort the near-field thermal profile. These conditions are indeed more restrictive, since they demand prior knowledge of the characteristic thermal decay length in the medium and an appropriate choice of the truncation distance.

+Neumann condition (insulation).

For the Neumann condition, the heat flux at the last node of the computational domain is forced to vanish, i.e., $\partial_r T(R_{sim}, t) = 0, \forall t$. This approach guarantees that no heat is allowed to leave the simulation region, effectively modelling an insulated outer boundary. Although this condition differs physically from the infinite-domain solution, it provides a complementary test case for assessing the stability and accuracy of the solver.

Figure S3 shows the numerical performance of the method for different particle diameters (D = 10, 40, 70, and 100 nm). In contrast with the Dirichlet case, both error metrics under the Neumann condition display a predominantly monotonic decay with increasing number of nodes. For instance, the global error decreases from values above 100% at $N_p$ = 5 down to below ~10% at $N_p$ ≥ 160 for the largest particles. Similarly, $\delta(T_{max})$ shows a systematic reduction, reaching errors below ~1% at the finest discretization.

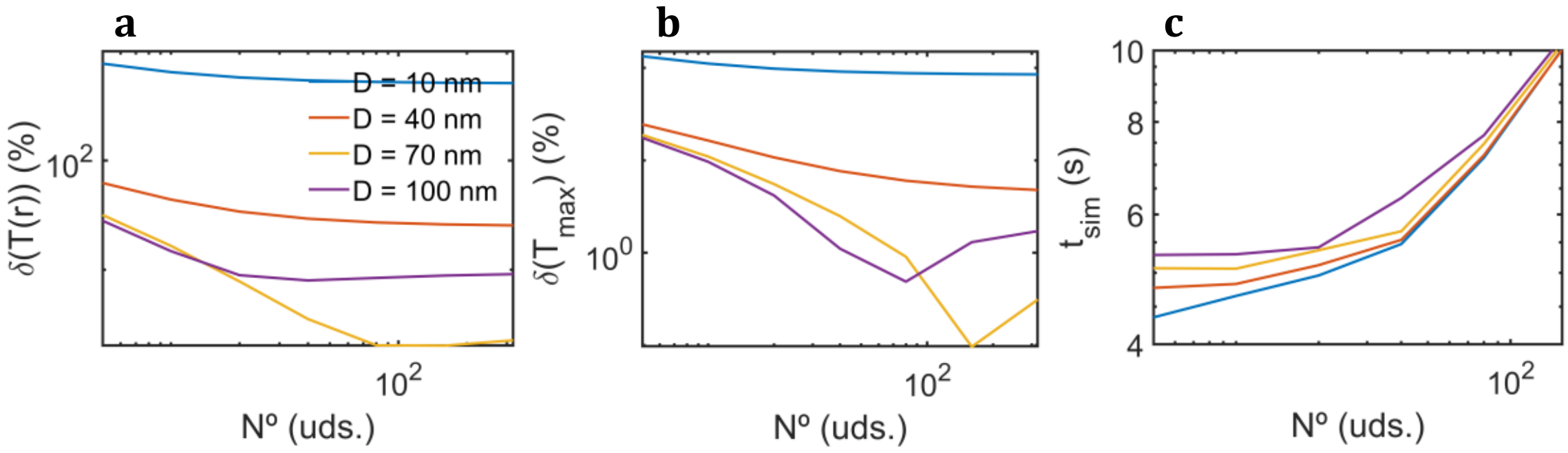

Figure S3. Relative error of the numerical solver compared to the analytical CW solution for different particle diameters ($D$ = 10, 40, 70, 100 nm). (a) Relative global error of the temperature profile $\delta(T(r))$, (b) error in the maximum temperature $\delta(T_{max})$, and (c) simulation time $t_{sim}$ as a function of the number of nodes inside the particle.

This behaviour reflects the fact that, under the Neumann boundary, the truncation of the domain does not artificially constrain the temperature profile to a fixed far-field value. Instead, the solution is free to adjust at the outer boundary, which reduces the bias observed in the Dirichlet case. However, the absence of a thermal reservoir also means that the overall temperature levels tend to be deviated compared to the infinite-domain solution, since no heat is extracted at the boundary. Interestingly, the configuration that shows the closest agreement with theory corresponds to D = 70 nm. For smaller particles, the simulation region is comparatively too limited, leading to heat accumulation and an overestimation of the thermal field. Conversely, for larger particles the external domain becomes disproportionately large, and the zero-flux boundary condition effectively pulls down the profile, producing an underestimation. In addition, the absorbed power itself increases with particle size, since it scales with both the particle volume and the absorption cross section. As a result, the thermal gradients at the boundary remain non-negligible even when the domain is large, and the impact of the Neumann condition persists. Thus, the optimal behaviour observed at intermediate diameters results from a balance between these competing effects.

Panel (c) confirms that the simulation time again grows approximately quadratically with $N_p$, from ~4–5 s at $N_p$ = 5 up to ~9–10 s at $N_p$ = 320. As in the Dirichlet case, the dependence on particle size is mild, with differences of less than ~1 s across the studied diameters. The computational scaling is thus primarily governed by the recursive solution of the linear systems, while the choice of boundary condition affects accuracy but not computational cost.

Overall, the Neumann condition yields more favourable convergence properties compared to Dirichlet, provided that the simulation domain is sufficiently large to minimize finite-size effects. Nevertheless, this advantage comes at the expense of an artificial thermal confinement that systematically shifts the numerical solution upward relative to the analytical infinite-domain case. Moreover, as shown in the analysis, achieving accurate results under Neumann also requires prior knowledge of the system behaviour in order to select an adequate simulation domain size, highlighting that the method is not entirely free from restrictive assumptions. In practice, this also implies that the simulation region may need to be chosen much larger than the structure itself, and it must grow even further for systems that are more efficient in generating heat. As a consequence, the memory requirements and the size of the matrices to be solved increase rapidly, which significantly impacts the computational cost.

+Robin condition (convective cooling).

For the Robin boundary condition, the heat flux across the particle–medium interface is governed by the interfacial thermal conductance $G$. This condition can be expressed as

$$-k_p \partial_r T_{r=a^-} = G(T(a^-) - T(a^+)) = -k_m \partial_r T_{r=a^+}$$

which generalizes the Neumann case: in the limit $G \to +\infty$, continuity of temperature is enforced at the interface, whereas for finite $G$ a thermal discontinuity develops that hinders heat transfer.

Figure S4 illustrates the numerical performance of the method for two representative cases: (a–c) $G \to +\infty$, corresponding to perfect continuity, and (d–f) a finite value $G$=200 MW m$^{-2}$ K$^{-1}$. For the ideal case $G \to +\infty$, both error metrics decrease monotonically with increasing discretization, reaching values below ~1% for both $\delta(T(r))$ and $\delta(T_{max})$ at the finest meshes. This behaviour demonstrates that the Robin condition with perfect interfacial conductance reproduces the analytical continuous-wave solution without introducing the systematic biases observed under Dirichlet or Neumann.

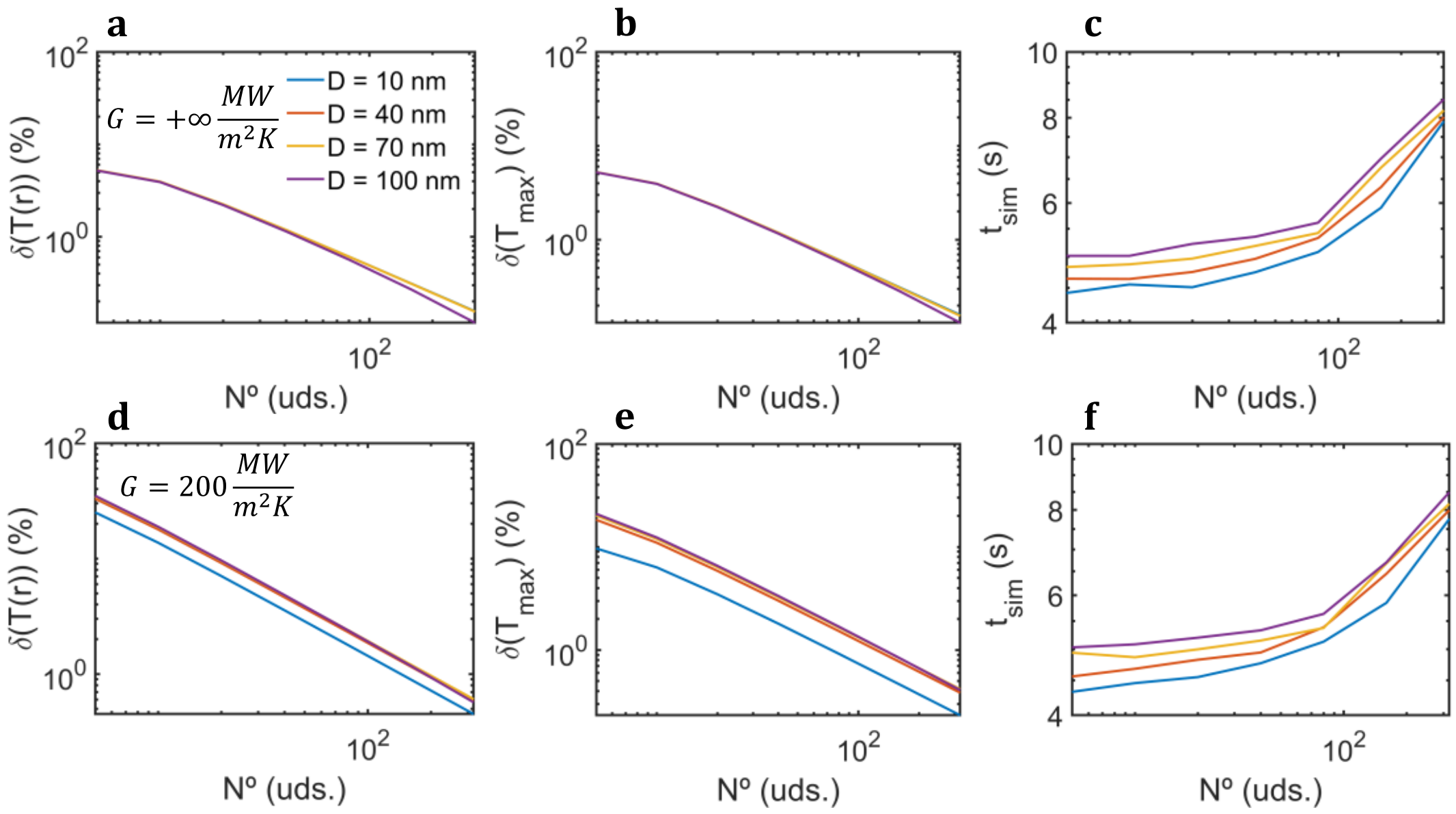

Figure S4. Relative error of the numerical solver compared to the analytical CW solution under Robin boundary conditions for different particle diameters (D = 10, 40, 70, 100 nm). Panels (a–c) correspond to perfect continuity at the interface ($G \to +\infty$), while panels (d–f) show the case of finite interfacial thermal conductance ($G$=200 MW m$^{-2}$ K$^{-1}$). (a,d) Relative global error of the temperature profile $\delta(T(r))$, (b,e) error in the maximum temperature $\delta(T_{max})$, and (c,f) simulation time $t_{sim}$ as a function of the number of nodes inside the particle.

When a finite interfacial conductance is introduced ($G$ = 200 MW m$^{-2}$ K$^{-1}$), the convergence trends remain monotonic, but the overall error levels are slightly higher across all particle sizes but still below 1% in all cases. This reflects the physical effect of interfacial resistance, which reduces heat transfer efficiency and enhances the thermal gradients across the particle–medium interface. Importantly, the numerical solver correctly captures this discontinuity, as evidenced by the consistent convergence towards the analytical Robin solution.

In both cases, the simulation times follow the same quadratic scaling with $N_p$ as observed for Dirichlet and Neumann, increasing from ~4–5 s to ~8–9 s. The dependence on particle size remains mild, confirming that computational cost is determined primarily by matrix size and solution strategy rather than by physical parameters such as particle diameter or interfacial conductance.

Overall, the Robin condition provides the most flexible and physically consistent framework, since it can interpolate between the limiting cases of perfect continuity and finite thermal resistance. This makes it the most appropriate boundary condition for modelling realistic nanoparticle–medium systems, where interfacial conductance plays a central role in defining the thermal response.

In conclusion, the developed software is capable of accurately, consistently, and efficiently computing the thermal response of spherical structures under continuous-wave illumination. By systematically benchmarking against the analytical solutions for Dirichlet, Neumann, and Robin boundary conditions, we have demonstrated that the solver reproduces both the global temperature profiles and the maximum temperature values with controllable errors. Moreover, the implementation remains computationally efficient, with simulation times scaling predictably with mesh refinement, thus providing a reliable framework for modelling photothermal processes in layered spherical nanoparticles.

SECTION 3: Supplementary calculations on thermal confinement in spherical nanoparticles.

To further support the results presented in the main text, we have performed an extended analysis of the thermal confinement in spherical metallic nanoparticles of different radii, using our in-house computational tool. The simulations covered particle radii ranging from 5 nm to 50 nm, in steps of 5 nm, and for each case the full thermoplasmonic response was computed under pulsed excitations with durations spanning from sub–electron–phonon coupling times up to the continuous wave (CW) limit. The temporal variable was subsequently normalized by the characteristic heating time of each structure, defined as

$$\tau_0^* = \frac{\tau_0}{\tau_d}$$

where $\tau_d = {a^2}/{\alpha}$ is characteristic diffusion time, $a$ is the particle radius and $\alpha$ is the thermal diffusivity of the surrounding medium. From the resulting spatial-temporal temperature field, the envelope of the temperature evolution was extracted. This envelope, which represents the maximum temperature reached at each point in space during the pulse, was normalized in order to remove the dependence on the total absorbed energy and thus isolate the effect of pulse duration. The analysis of this normalized envelope allows us to construct the maps shown in Figure S5, which display the normalized confinement radius $r_\gamma$ (defined as the radial position where the envelope decays by a factor $\gamma$) as a function of the pulse duration and the nanoparticle radius. Importantly, the confinement radius was normalized by the particle radius itself, making it a direct indicator of how spatially confined the thermal field remains around the structure.

Thus, Figure S5 presents a rescaled version of Figure 1 of the main text. In this version, the vertical axis corresponds to the logarithm of the normalized pulse duration, $\log_{10}\tau_0^*$, while the horizontal axis shows the particle radius. The colour scale represents $r_\gamma$ and the contour lines indicate constant confinement factors for different $\gamma$ values. Remarkably, once the excitation time is normalized, the contour lines become almost perfectly horizontal across all panels, evidencing that the degree of thermal confinement is independent of the particle size. This invariance implies that a single-particle simulation can capture the general behaviour of the system, provided that the results are properly rescaled in terms of $\tau_d$. This observation dramatically reduces the computational load while preserving the physical generality of the conclusions.

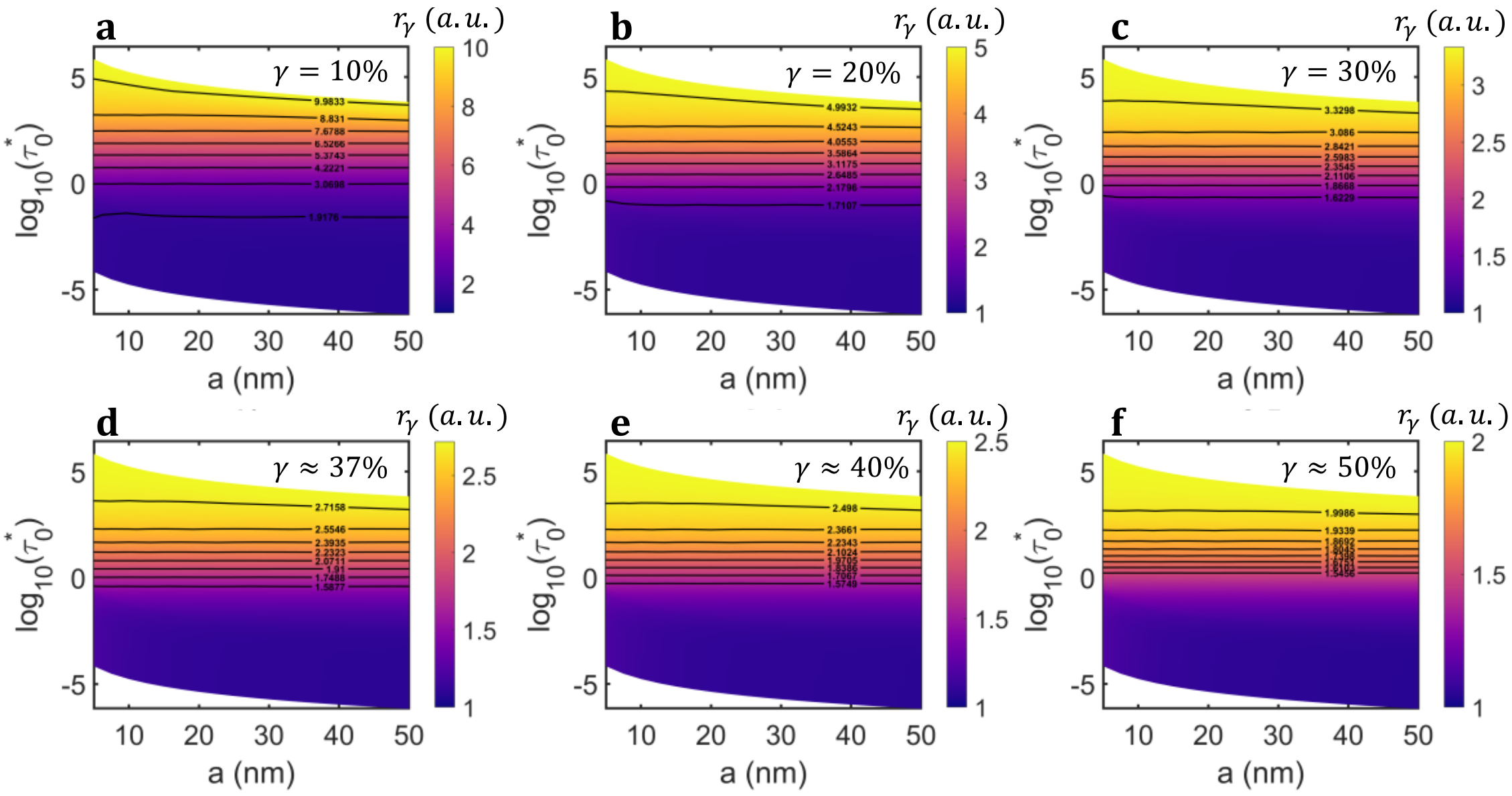

Figure S5. Mapping of the normalized thermal confinement radius ($r_\gamma$) as a function of the normalized pulse duration ($\tau_0^*$) and the particle radius ($a$) in the case of spherical gold nanoparticles. Panels a-f correspond to different confinement levels $\gamma = 0.1, 0.2, 0.3, e^{-1}, 0.4$ and 0.5 respectively.

Across all panels a-f (from $\gamma = 10\%$ to $\gamma = 50\%$), the same qualitative behaviour is observed: as the normalized excitation time increases, the confinement gradually decreases, meaning that the thermal energy extends further into the surrounding medium. In the ultrafast regime (lower region of each map), the confinement is stronger, with $r_\gamma$ values close to unity, indicating that the temperature field is essentially restricted to the nanoparticle volume. As the excitation duration increases, $r_\gamma$ grows progressively, reflecting enhanced diffusion. For instance, in the case $\gamma = 10\%$ (panel a), the confinement radius increases from $r_\gamma \approx 1.9a$ at short pulses up to $r_\gamma \approx 10a$ in the CW regime, where the system reaches the steady-state temperature distribution. Notably, the upper limit of each colourbar corresponds precisely to this CW value, which can be analytically predicted from the stationary heat equation. The agreement between the numerical and analytical results further validates the accuracy of the developed solver.

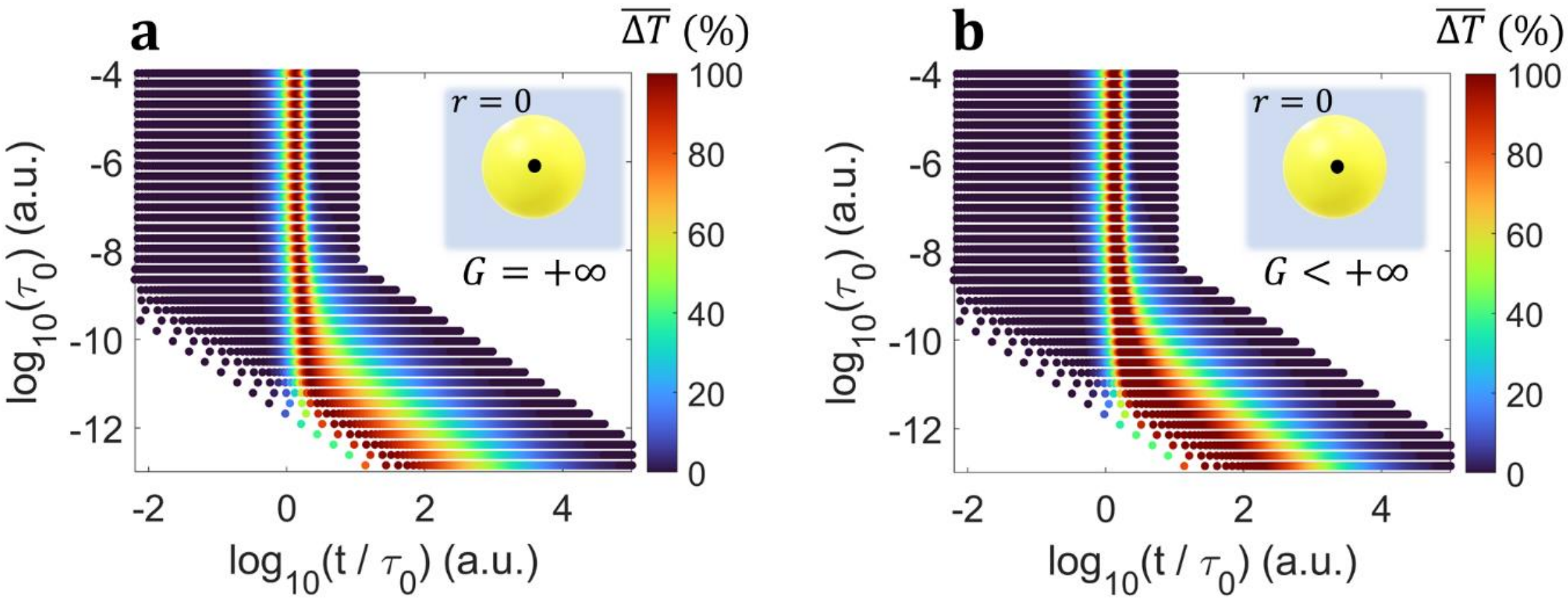

Figure S6. Normalized temperature increase in the centre of the nanoparticle for the cases of (a) infinite ITC and (b) finite ITC.

Figure S6 extends the analysis of Figure 3a–b by examining the normalized temperature dynamics inside the nanoparticle, at $r = 0$. Using the same normalization procedures, the maps show that for long pulses both the infinite- and finite-ITC cases converge, reflecting that heat has enough time to diffuse across the interface and the internal response becomes effectively independent of $G$. For shorter pulses, however, clear differences emerge: the finite-ITC case exhibits a sharper and more compressed temperature rise, indicating stronger thermal accumulation inside the particle when heat cannot escape efficiently. This behaviour is opposite to what occurs in the surrounding medium and highlights the dual role of the interface: it suppresses heat extraction externally while enhancing thermal confinement internally. Figure S6 shows that the influence of ITC is not limited to the temperature outside the nanoparticle but also significantly modifies the transient thermal response within the particle, especially under ultrafast excitation.